\documentclass[acmlarge,authorversion]{acmart}

\usepackage[ruled]{algorithm2e} 

\SetAlFnt{\small}
\SetAlCapFnt{\small}
\SetAlCapNameFnt{\small}
\SetAlCapHSkip{0pt}

\usepackage{subfigure}
\usepackage{caption}
\usepackage{algorithmic}
\usepackage{colortbl}
\usepackage{subfigure}
\usepackage{graphicx}
\usepackage{hyperref}
\usepackage{placeins}

\usepackage{siunitx}
\usepackage{enumitem}
\usepackage{listings}

\usepackage{amsmath,amsfonts,bm}

\def\eqref#1{equation~\ref{#1}}

\def\1{\bm{1}}

\def\rva{{\mathbf{a}}}

\def\rvc{{\mathbf{c}}}

\def\rvo{{\mathbf{o}}}

\def\rvq{{\mathbf{q}}}

\def\rvs{{\mathbf{s}}}

\def\rvx{{\mathbf{x}}}
\def\rvy{{\mathbf{y}}}

\DeclareMathAlphabet{\mathsfit}{\encodingdefault}{\sfdefault}{m}{sl}
\SetMathAlphabet{\mathsfit}{bold}{\encodingdefault}{\sfdefault}{bx}{n}

\newcommand{\expec}{\mathbb{E}}

\newcommand{\code}[1]{\colorbox[rgb]{0.9,0.9,0.9}{\texttt{#1}}}

\begin{document}

\title{MimicKit: A Reinforcement Learning Framework for Motion Imitation and Control}

\author{Xue Bin Peng}
\email{xbpeng@sfu.ca}
\orcid{0002-3677-5655}
\affiliation{
  \institution{Simon Fraser University}
}
\affiliation{
  \institution{NVIDIA}
}

\begin{abstract}
\textbf{MimicKit} is an open-source framework for training motion controllers using motion imitation and reinforcement learning. The codebase provides implementations of commonly-used motion-imitation techniques and RL algorithms. This framework is intended to support research and applications in computer graphics and robotics by providing a unified training framework, along with standardized environment, agent, and data structures. The codebase is designed to be modular and easily configurable, enabling convenient modification and extension to new characters and tasks. The open-source codebase is available at: \href{https://github.com/xbpeng/MimicKit}{\texttt{https://github.com/xbpeng/MimicKit}}.
\end{abstract}

\begin{teaserfigure}
  \includegraphics[width=1\textwidth]{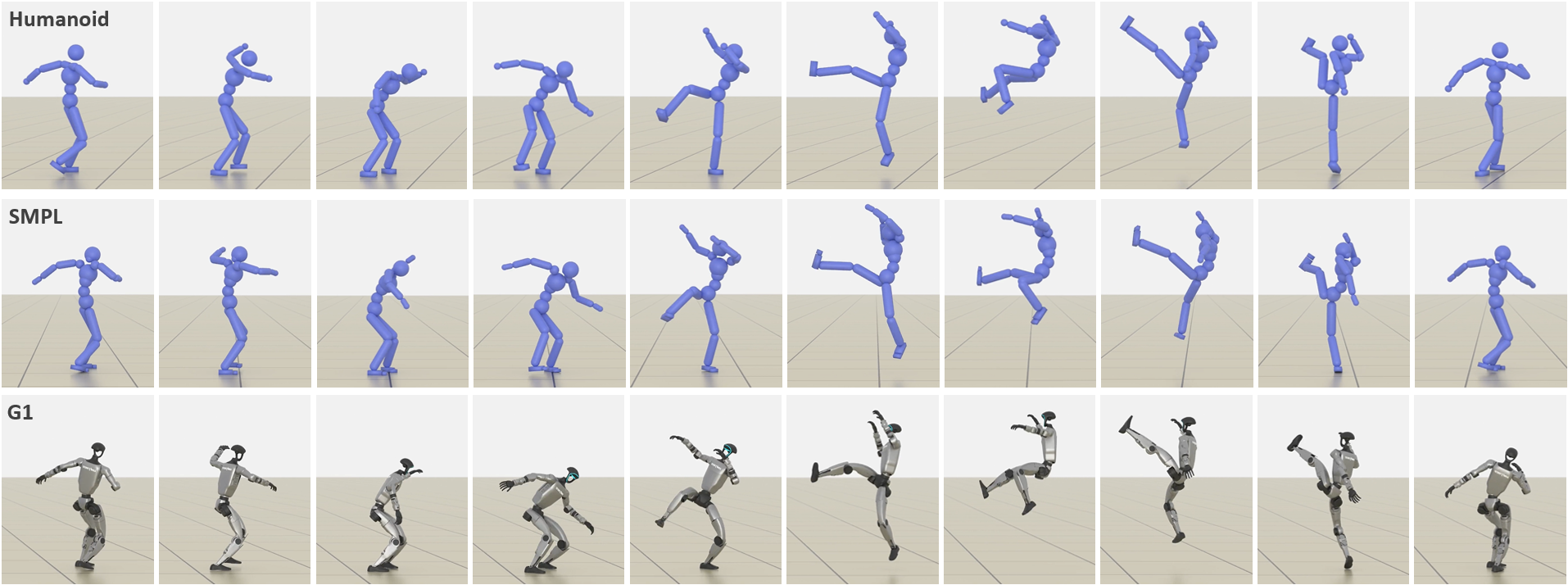}
  \caption{MimicKit provides a suite motion imitation methods that can be used to train diverse simulated agents to perform highly dynamic and life-like motor skills. In this example, a variety of physically simulated humanoid characters are trained to perform a spinkick motion.}
  \label{fig:teaser}
\end{teaserfigure}

\maketitle

\section{Introduction}
Reinforcement-learning (RL) based motion imitation techniques have become a versatile and effective paradigm for constructing motion controllers that are able to produce agile, life-like behaviors for both simulated characters and robots in the real world. Although the many of the core ideas are conceptually simple, building effective motion imitation systems requires careful attention to numerous nuances and detailed design decisions that are often challenging to implement in practice. MimicKit is designed to lower the barrier for experimentation and reproducible research in this field by bringing together a suite of high-quality implementations of training methods and tools into a single unified and extensible framework.

\section{Background}
In MimicKit, most models are trained using reinforcement learning, where an agent interacts with an environment according to a policy $\pi$ in order to optimize a given objective \cite{Sutton2018}. At each time step $t$, the agent receives an observations $\rvo_t$ of the environment, which provides partial information of the state $\rvs_t$ of the underlying system. The agent responds by sampling an action from a policy $\rva_t \sim \pi(\rva_t | \rvo_t)$. The agent then executes the action, which leads to a new state $\rvs_{t+1}$, sampled according to the dynamics of the environment $\rvs_{t+1} \sim p(\rvs_{t+1} | \rvs_t, \rva_t)$. The agent in turn receives a scalar reward $r_t = r(\rvs_t, \rva_t, \rvs_{t+1})$, and a new observation $\rvo_{t+1}$ of the next state $\rvs_{t+1}$. The agent's objective is to learn a policy that maximizes its expected discounted return $J(\pi)$,
\begin{equation}
    J(\pi) = \expec_{p(\tau | \pi)} \left[ \sum_{t=0}^{T-1} \gamma^t r_t \right],
\label{eqn:rl_objective}
\end{equation}
where $p(\tau | \pi)$ represents the likelihood of a trajectory $\tau = \{\rvo_0, \rva_0, r_0, \rvo_1, ..., \rvo_{T-1}, \rva_{T-1}, r_{T-1}, \rvo_T \}$ under $\pi$. $T$ denotes the time horizon of a trajectory, and $\gamma \in [0, 1]$ is a discount factor. Each trajectory corresponds to one \emph{episode} of interactions between the agent and the environment.

\begin{figure}[t]
\centering
\includegraphics[width=0.6\textwidth]{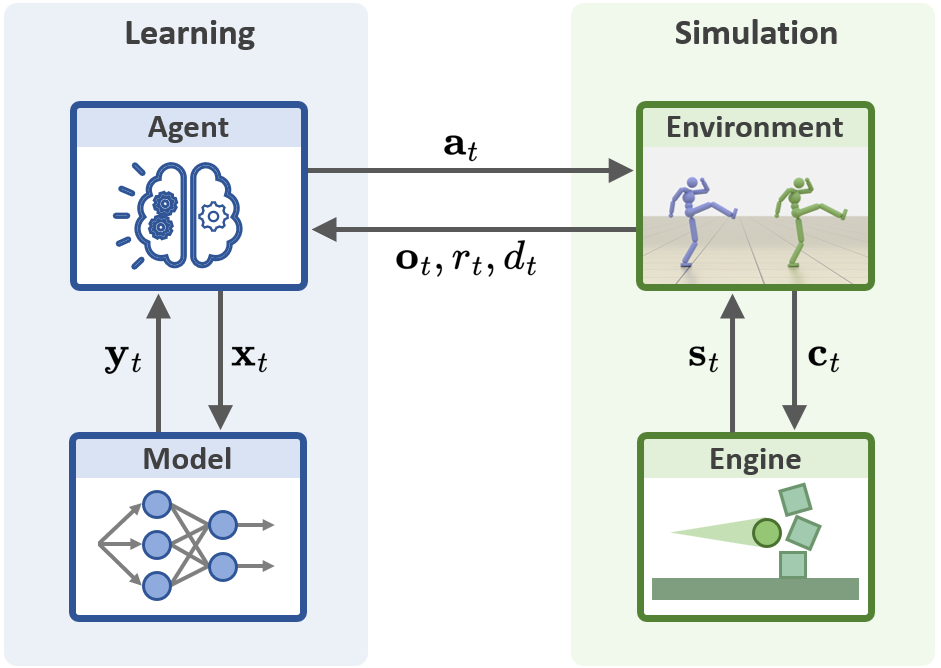}
\caption{Schematic overview of the MimicKit framework. The main components of the system are 1) the \texttt{Agent}, 2) the \texttt{Model}, 3) the \texttt{Environment}, and 4) the \texttt{Engine}. The learning algorithms are implemented primarily through the \texttt{Agent} and \texttt{Model}, while the \texttt{Environment} and \texttt{Engine} are responsible for simulating the desired task.}
\label{fig:overview}
\end{figure}

\section{System Overview}
A schematic overview of the MimicKit framework is provided in Figure~\ref{fig:overview}. The core components of MimicKit consist of: 1) the \texttt{Agent}, 2) the \texttt{Model}, 3) the \texttt{Environment}, and 4) the \texttt{Engine}. The learning algorithms are implemented primarily through the \texttt{Agent} and the \texttt{Model}, while the \texttt{Environment} and \texttt{Engine} are responsible for simulating the desired task. These components are designed to be modular and composable, enabling users to combine different learning algorithms, model architectures, characters, tasks, and simulators. The simulations are implemented using vectorized environments, which can be massively parallelized using GPU simulators for high-throughput data collection during training. The environments and learning algorithms are designed to be character-agnostic, enabling the overall system to be easily configured to support characters with different morphologies, including humanoid and non-humanoid characters, such as quadrupedal robots.

\subsection{Agent} 
The \texttt{Agent} class is responsible for implementing the learning algorithm and managing data recorded through interactions with the environment. Implementations for a suite of different agents are provided in \code{mimickit/learning/}. At each timestep $t$, the \texttt{Agent} receives an observation $\rvo_t$ from the \texttt{Environment}. This observations is processed into an input $\rvx_t$ for the \texttt{Model}, which can include pre-processing steps such as observation normalization. The \texttt{Model} is then queried with the processed input $\rvx_t$, which produces an output $\rvy_t$. The \texttt{Model} outputs $\rvy_t$ may specify parameters of an action distribution, value function predictions, discriminator predictions, or other quantities required by the learning algorithm. The \texttt{Agent} then extracts an action $\rva_t$ from the model outputs, and applies $\rva_t$ to the \texttt{Environment}. The \texttt{Environment} in turn transitions to a new state $\rvs_{t+1}$ and provides the \texttt{Agent} with the next observations $\rvo_{t+1}$, reward $r_t$, and a done flag $d_t$. The done flag $d_t$ indicates if the current episode has been terminated. 

In each iteration, the \texttt{Agent} repeats this interaction loop with the \texttt{Environment} until a designated number of timesteps has been collected. The data collected through these interactions are stored in an experience buffer implemented in \code{mimickit/learning/experience\_buffer.py}. Once a sufficiently large batch of data has been collected, the \texttt{Agent} then uses the data to update its \texttt{Model}. The agent configuration files, located in \code{data/agents/}, are used to specify the type of \texttt{Agent} to use for training, as well as its associated hyperparameters.

\subsection{Model} 
While the \texttt{Agent} implements the learning procedure, the \texttt{Model} is responsible for implementing the underlying neural network architecture used in the learning process. Each \texttt{Agent} is paired with a corresponding \texttt{Model}, located in \code{mimickit/learning/}. For an actor-critic algorithm, such as PPO \citep{PPO2017}, the model may contain multiple neural networks, one for the policy (i.e. actor) and the value function (i.e. critic). For methods such as AMP, an additional network might be constructed for the discriminator. The \texttt{Agent} can query the \texttt{Model}'s various networks with the appropriate input $\rvx_t$, and the \texttt{Model} returns the corresponding output $\rvy_t$. The \texttt{Model}'s network architectures are specified through the \code{model} field in the agent configuration file.

\subsection{Environment}
The \texttt{Environment} implements the task-specific logic necessary to simulated a desired task. This class is used to define the interface through which the agent observes and interacts with its surrounding environment. At each timestep $t$, the \texttt{Environment} constructs the observation $\rvo_t$ based on the state $\rvs_t$ of the world determined by the \texttt{Engine}. The observation $\rvo_t$ can contain proprioceptive information on the character's body, information on the configuration of surrounding objects, as well as task-specific information, such as the target locations and steering commands.

Upon receiving the action $\rva_t$ from the \texttt{Agent}, the \texttt{Environment} processes $\rva_t$ into a command $\rvc_t$. The command is then applied to the \texttt{Engine} to update the state of the underlying system, which can be modeled by a simulator or correspond to a real-world system. The environment update is performed through a step function:
\begin{lstlisting}[basicstyle=\ttfamily\small,backgroundcolor=\color{gray!20!white}]
    obs, r, done, info = self._env.step(action)
\end{lstlisting}
The step function returns a new observation $\rvo_{t+1}$ and a reward $r_t$ for the state transition. Furthermore, an \code{info} dictionary can be used to store additional information from the \texttt{Environment}, such as auxiliary observations for a critic or discriminator. Finally, the done flag $d_t$ indicates if the current episode has terminated. The done flag can assume 4 different values, as defined in \code{mimickit/envs/base\_env.py}, depending on the conditions under which an episode was terminated. The different done flags include:
\begin{itemize}
    \item \code{NULL}: The episode has not been terminated.
    \item \code{FAIL}: The episode terminated due to a failure, such as the character falling down. This flag can be used to apply a terminal penalty when calculating returns during training.
    \item \code{SUCC}: The episode terminated due to successfully completing a task, such as the character successfully reaching the target location. This flag can be used to apply a terminal bonus when calculating returns during training.
    \item \code{TIME}: The episode is terminated due to a time limit, but should in principle continue after the last timestep of the episode. In the event that a trajectory is truncated due to time, bootstrapping with a value function can be used to estimate future returns, as if the trajectory had continued after the last timestep. This enables the learning algorithms to emulate infinite-horizon MDPs given finite-length trajectories.
\end{itemize}
The configuration of the \texttt{Environment} is specified through environment configuration files located in \code{data/envs/}.

\subsection{Engine} 
While the \texttt{Environment} implements the high-level logic for simulating a particular task, the low-level simulation of the world is delegated to an \texttt{Engine}. The \texttt{Engine} class, implemented in
\code{mimickit/engines/engine.py}, provides a unified API that abstracts away the low-level details of how an \texttt{Environment} is simulated. Different \texttt{Engines} can be constructed for different physics engines and real-world robotic systems. This enables a specific task and environment to be instantiated through different underlying simulators and physical robots. MimicKit currently supports Isaac Gym \citep{IsaacGym2021}, Isaac Lab \citep{IsaaccLab2025}, and Newton \citep{newton}. Additional \texttt{Engines} will be introduced in the future to support other physics simulators, as well as deployment on real robots.

At each timestep, the \texttt{Engine} receives the command $\rvc_t$ from the \texttt{Environment}, and returns an updated state $\rvs_{t+1}$. The representation of $\rvc_t$ depends on the control modes that are supported by a specific \texttt{Engine}. For example, the IsaacGym \texttt{Engine}, implemented in \code{mimickit/engines/isaac\_gym\_engine.py} supports the following control modes:
\vspace{0.2cm}
\begin{itemize}
    \item \code{none}: Commands have no effect on the simulation. This mode can be useful for visualization and debugging.
    \item \code{pos}: Commands specify target rotations for PD controllers, which support both 1D revolute joints and 3D spherical joints.
    \item \code{vel}: Commands specify target velocities for each joint.
    \item \code{torque}: Commands directly specify torques for each joint.
    \item \code{pd\_1d}: Commands specify target rotations for 1D revolute joints. This control mode can only be applied to morphologies that solely consist of 1D revolute joints, and does not support 3D spherical joints. This control mode is best suited for simulating robots that only contain 1D revolute joints.
\end{itemize}
\vspace{0.2cm}
The configuration of the \texttt{engine} can be specified through the \code{engine} field in the environment configuration file. The environment file can be used to specify the type of \texttt{Engine} to use for simulation, along with parameters such as the control model, control frequency, simulation frequency, etc.

\clearpage

\section{Methods}
MimicKit provides a suite of motion imitation methods for training controllers. These methods offer different characteristics and trade-offs, and an appropriate method should be selected based on the requirements of a target application. Example argument files are provided in the arguments directory \code{args/} for using the various methods.
\vspace{0.2cm}

\subsection{DeepMimic}
\begin{figure}[h]
\centering
\includegraphics[width=1\textwidth]{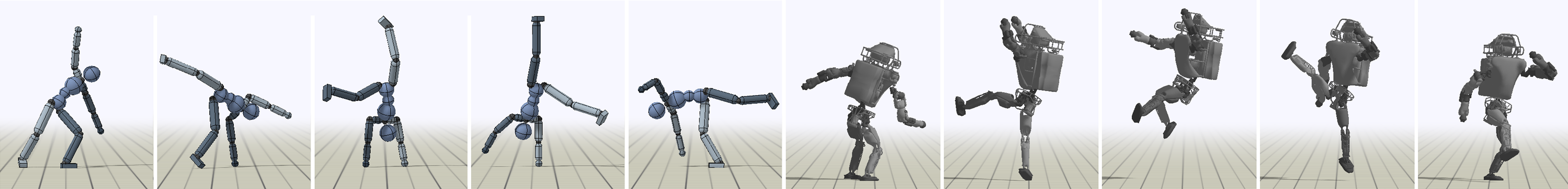}
\label{fig:deepmimic}
\end{figure}
\vspace{-0.6cm}
DeepMimic is a simple RL-based motion tracking method \citep{2018-TOG-deepMimic}, which trains a tracking controller to follow target reference motions. This method is very general and reliable, and has been successfully applied to train controllers for a wide range of behaviors. DeepMimic is often a good starting point before considering more sophisticated techniques, and can be a highly effective method for applications that require precise replication of a target reference motion. However, a key limitation of DeepMimic is that the motion tracking objective used during training often leads to inflexible policies that are restricted to closely following a given reference motion. This can limit the agent's ability to modify and adapt behaviors in the dataset as necessary to perform new tasks. 

Example arguments for running DeepMimic are provided in \code{args/deepmimic\_humanoid\_ppo\_args.txt}. The reference motion data used for training can be specified using the \code{motion\_file} in the environment configuration file \code{data/envs/deepmimic\_humanoid\_env.yaml}.
\vspace{0.2cm}

\subsection{Adversarial Motion Priors (AMP)}
\begin{figure}[h]
\centering
\includegraphics[width=1\textwidth]{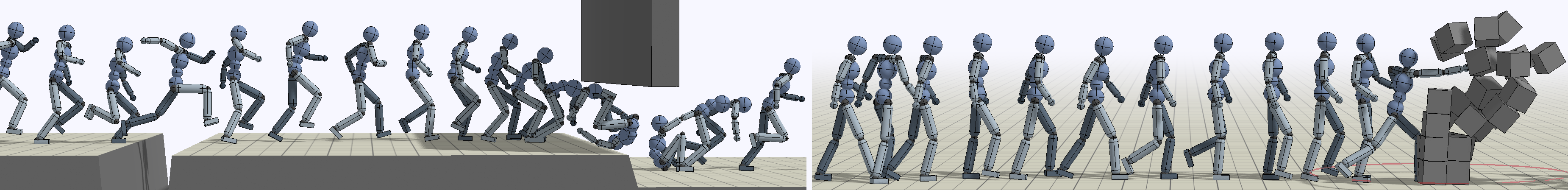}
\label{fig:AMP}
\end{figure}
\vspace{-0.6cm}
Unlike DeepMimic, which trains a controllers to closely track a given reference motion, AMP is an adversarial distribution-matching method that aims to imitate the overall behavioral distribution (i.e. style) depicted in a dataset of motion clips \citep{2021-TOG-AMP}, without explicitly tracking any specific motion clip. AMP provides more versatility than tracking-based methods, providing the agent with more flexibility to compose and adapt behaviors in the dataset in order to perform new tasks. However, a key drawback of distribution-matching methods, such as AMP, is that they are more prone to converging to local optima, especially for challenging, highly dynamics motions. Therefore AMP may struggle more to closely replicate challenging behaviors, compared to tracking-based methods, such as DeepMimic. 

Example arguments for using AMP to imitate individual motion clips, without auxiliary tasks, are provided in \code{args/amp\_humanoid\_args.txt}. An example for training an AMP model with auxiliary tasks is provided in \code{args/amp\_location\_humanoid\_args.txt}.

\clearpage

\subsection{Adversarial Skill Embeddings (ASE)} 
\begin{figure}[h]
\centering
\includegraphics[width=1\textwidth]{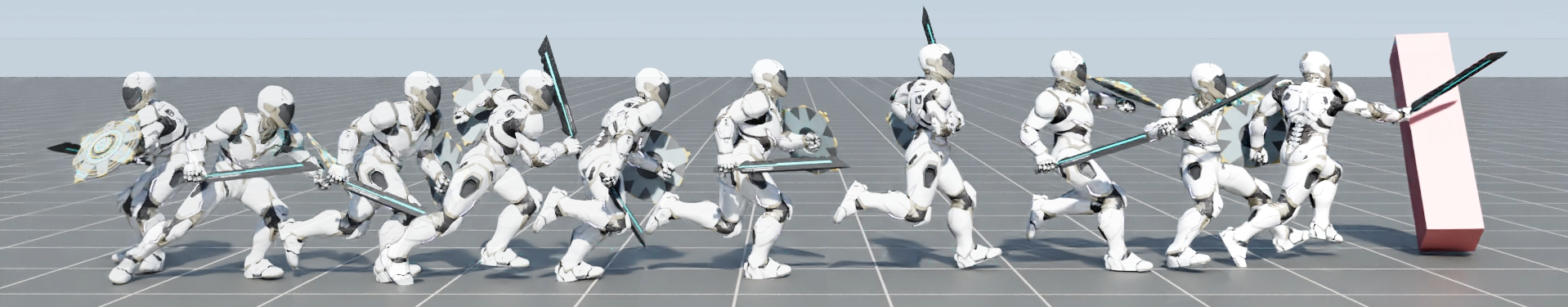}
\label{fig:ASE}
\end{figure}
\vspace{-0.6cm}
ASE is an adversarial methods for training reusable generative controllers \citep{2022-TOG-ASE}. This method combines adversarial imitation learning with a mutual information-based skill discovery objective to learn latent skill embeddings. Points in the latent space can be mapped to diverse behaviors by the ASE controller. Once trained, the ASE controller can be reused to perform new tasks by training task-specific high-level controllers to select skills from the learned latent space. Example arguments for training ASE models are provided in \code{args/ase\_humanoid\_args.txt}.
\vspace{0.2cm}

\subsection{Adversarial Differential Discriminator (ADD)} 
\begin{figure}[h]
\centering
\includegraphics[width=1\textwidth]{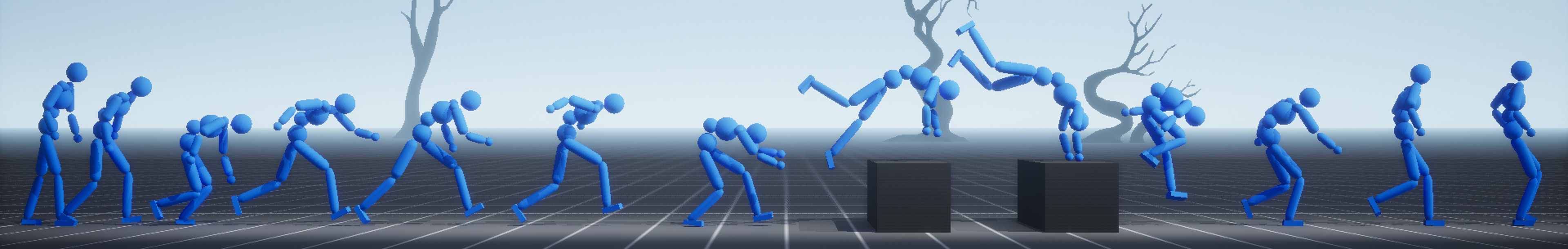}
\label{fig:ADD}
\end{figure}
\vspace{-0.6cm}
ADD is an adversarial motion tracking method that uses a differential discriminator to automatically learn adaptive motion tracking objectives \citep{zhang2025ADD}. This method can mitigate the manual effort required to design and tune tracking reward functions for different characters and motions. Example arguments for training ADD models are provided in \code{args/add\_humanoid\_args.txt}.

\hfill\newline
Most of the methods in MimicKit are implemented using proximal policy optimization (PPO) as the underlying RL algorithm \citep{PPO2017}. PPO is currently the most commonly-used RL algorithm for motion control tasks, and can be effectively scaled with high-throughput GPU simulators. However, since PPO is an on-policy algorithm \cite{Sutton2018}, it can be notoriously sample inefficient. Our framework provides an off-policy algorithm, Advantage-Weighted Regression (AWR) \citep{AWRPeng19}, as an alternative to PPO for settings that may require off-policy RL algorithms.

\section{Instructions}

In this section, we provide starter instructions for installing MimicKit, training models, testing models, and an overview of basic tools to assist in common workflows.

\subsection{Installation}
MimicKit supports different simulator backends. First, install the simulator of your choice from their respective source. The simulators that are currently supported include: Isaac Gym \citep{IsaacGym2021}, Isaac Lab \citep{IsaaccLab2025}, and Newton \citep{newton}. MimicKit can then be installed by following the steps below:
\begin{enumerate}
    \item Install the dependencies from \code{requirements.txt}:
    \begin{lstlisting}[basicstyle=\ttfamily\small,backgroundcolor=\color{gray!20!white}]
    pip install -r requirements.txt
    \end{lstlisting}

    \item Download assets and motion data from the \href{https://1sfu-my.sharepoint.com/:u:/g/personal/xbpeng_sfu_ca/EclKq9pwdOBAl-17SogfMW0Bved4sodZBQ_5eZCiz9O--w?e=bqXBaa}{\texttt{data repository}}, then extract the contents into the data directory \code{data/}.
\end{enumerate}
After completing these steps, MimicKit should be ready for use.

\subsection{Training}
To train a model, a typical training command will be as follows:
\begin{lstlisting}[basicstyle=\ttfamily\small,backgroundcolor=\color{gray!20!white}]
python mimickit/run.py --mode train --num_envs 4096 \
  --engine_config data/engines/isaac_gym_engine.yaml \
  --env_config data/envs/deepmimic_humanoid_env.yaml \
  --agent_config data/agents/deepmimic_humanoid_ppo_agent.yaml \
  --visualize true \
  --out_dir output/
\end{lstlisting}
The arguments consist of
\begin{itemize}
    \item \code{-{}-mode} selects either \code{train} or \code{test} mode.
    \item \code{-{}-num\_envs} specifies the number of parallel environments used for simulation.
    \item \code{-{}-engine\_config} specifies the configuration file for the Engine to select different simulator backends.
    \item \code{-{}-env\_config} specifies the configuration file for the environment.
    \item \code{-{}-agent\_config} specifies configuration file for the agent.
    \item \code{-{}-visualize} enables visualization. Rendering should be disabled for faster training.
    \item \code{-{}-out\_dir} specifies the output directory where the models and logs will be saved.
    \item \code{-{}-logger} specifies the logger used to record training statistics. The options are TensorBoard \code{tb} or \code{wandb}.
\end{itemize}
Instead of specifying all arguments through the command line, arguments can also be loaded from an argument file \code{arg\_file}:
\begin{lstlisting}[basicstyle=\ttfamily\small,backgroundcolor=\color{gray!20!white}]
python mimickit/run.py --arg_file args/deepmimic_humanoid_ppo_args.txt
\end{lstlisting}
The arguments in \code{arg\_file} are treated the same as command line arguments. When using an argument file, additional command line arguments can be included to override the arguments in the \code{arg\_file}. A library of arguments are provided in the arguments directory \code{args/} for training models and using various tools.

\subsection{Distributed Training}
The standard training command will train a model using a single process. To accelerate training, distributed training with multi-CPU or multi-GPU can be used with the following command:
\begin{lstlisting}[basicstyle=\ttfamily\small,backgroundcolor=\color{gray!20!white}]
python mimickit/run.py --arg_file args/deepmimic_humanoid_ppo_args.txt \
   --devices cuda:0 cuda:1
\end{lstlisting}
where \code{-{}-devices} specifies the devices used for training, which can be \code{cpu} or \code{cuda:\{i\}}. Multiple devices can be provided to parallelize training across multiple processes.

\subsection{Testing}
During training, the latest model parameters will be saved to a checkpoint \code{.pt} file, specified by \code{-{}-out\_model\_file}. A typical command to test a trained model will be as follows:
\begin{lstlisting}[basicstyle=\ttfamily\small,backgroundcolor=\color{gray!20!white}]
python rl_forge/run.py --arg_file args/deepmimic_humanoid_ppo_args.txt \
  --num_envs 4 \
  --visualize true \
  --mode test \
  --model_file data/models/deepmimic_humanoid_spinkick_model.pt
\end{lstlisting}
\code{-{}-mode test} specifies that the code should be run in testing mode. \code{-{}-model\_file} specifies the \code{.pt} file that contains the parameters of the trained model. Pretrained models are provided in \code{data/models/}, and the corresponding training log files are available in \code{data/logs/}.

\subsection{Visualizer User Interface}

A simple interface is provided for controlling the visualizer, which is shared across different Engines.
\begin{itemize}
    \item \textbf{Camera control:} Hold \code{Alt} key and drag with the left mouse button to pan the camera. Scroll with the mouse wheel to zoom in/out.
    \item \textbf{Pause Simulation:} \code{Enter} key can be used to pause/unpause the simulation
    \item \textbf{Step Simulation:} \code{Space} key can be used to step the simulator one step at a time.
\end{itemize}
Each Engine may also implement different additional controls depending on the underlying simulator.

\subsection{Visualizing Training Logs}
\begin{figure}[t]
	\centering
    \subfigure{\includegraphics[height=0.17\linewidth]{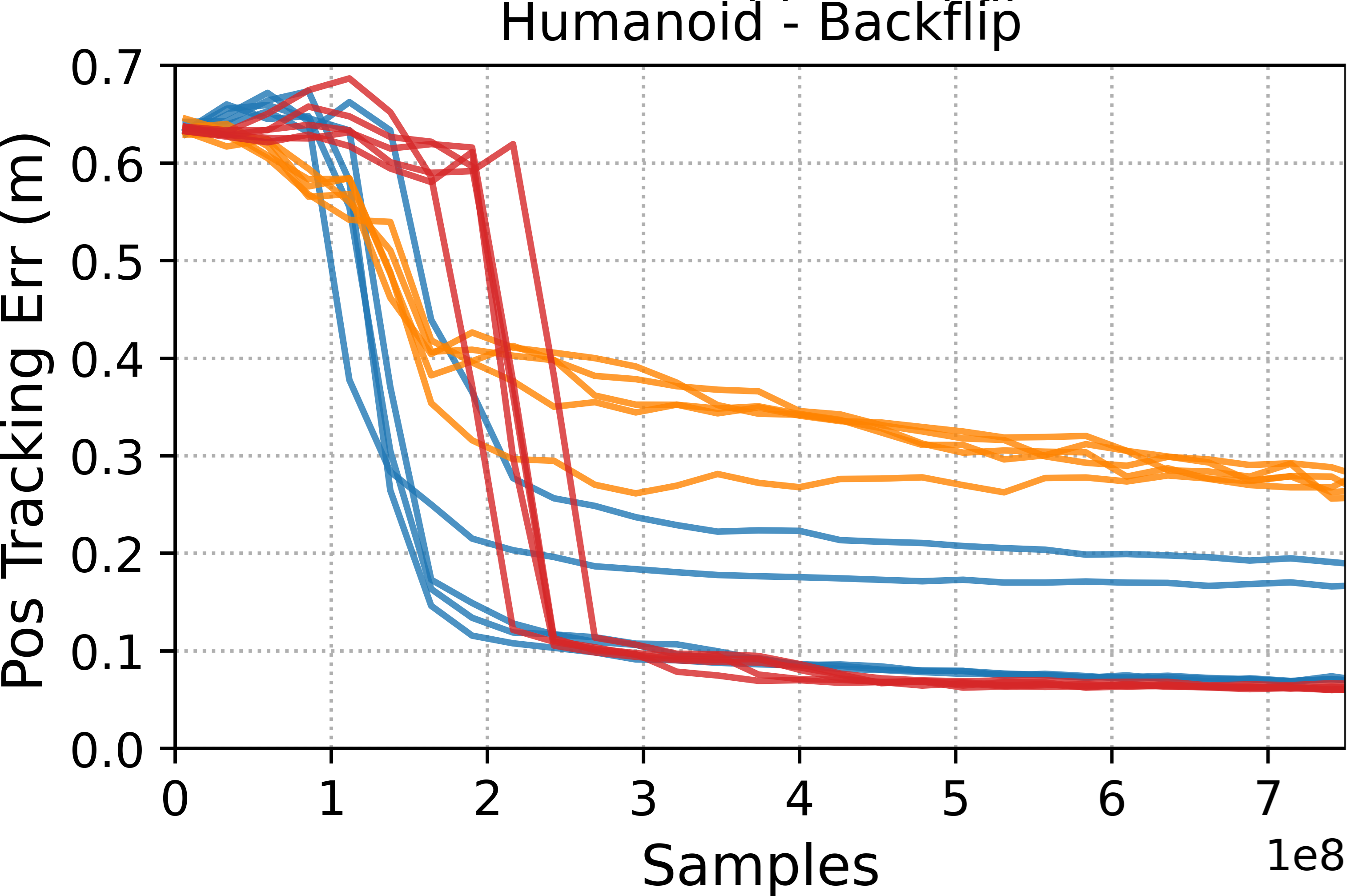}}
    \subfigure{\includegraphics[height=0.17\linewidth]{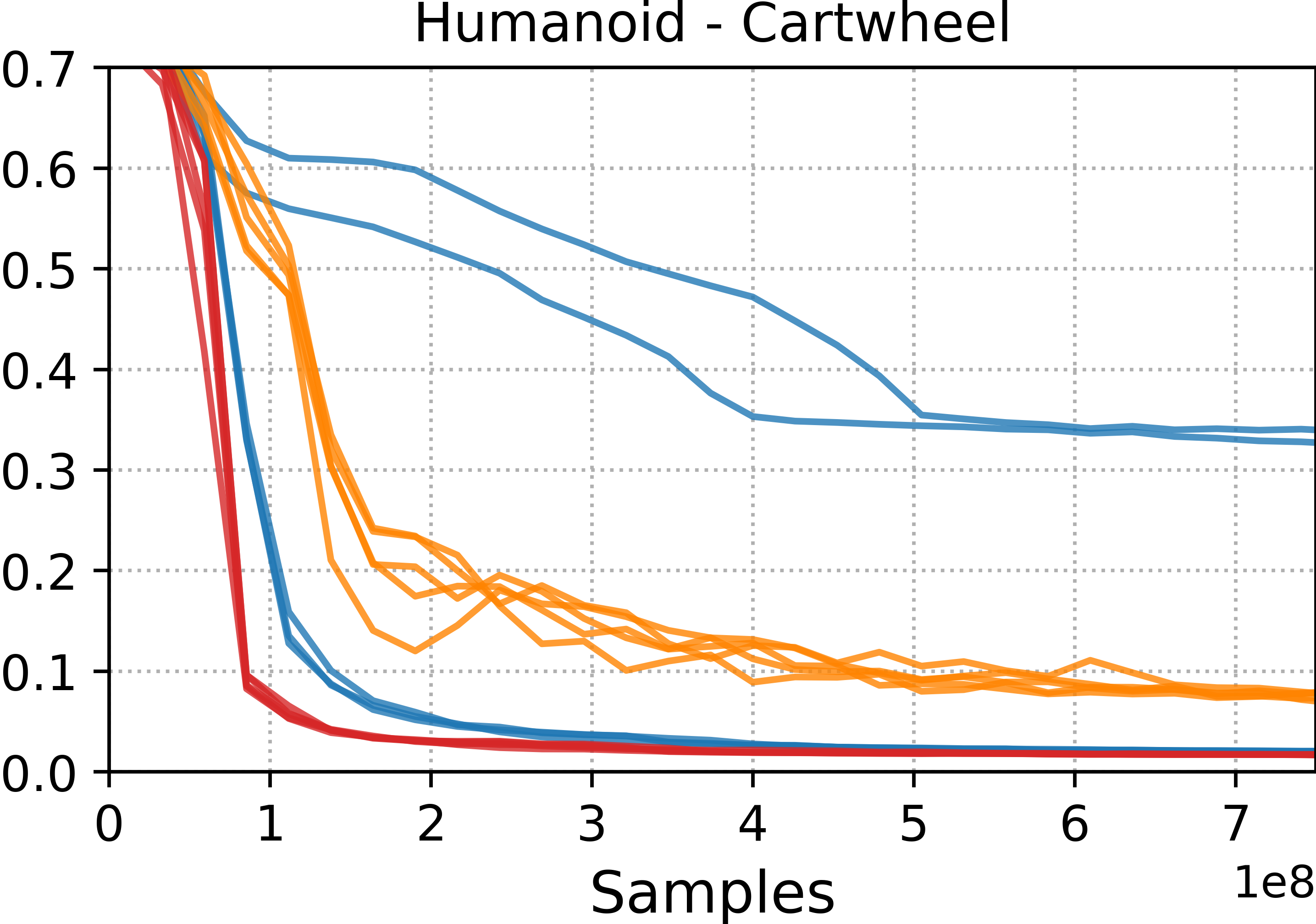}}
    \subfigure{\includegraphics[height=0.17\linewidth]{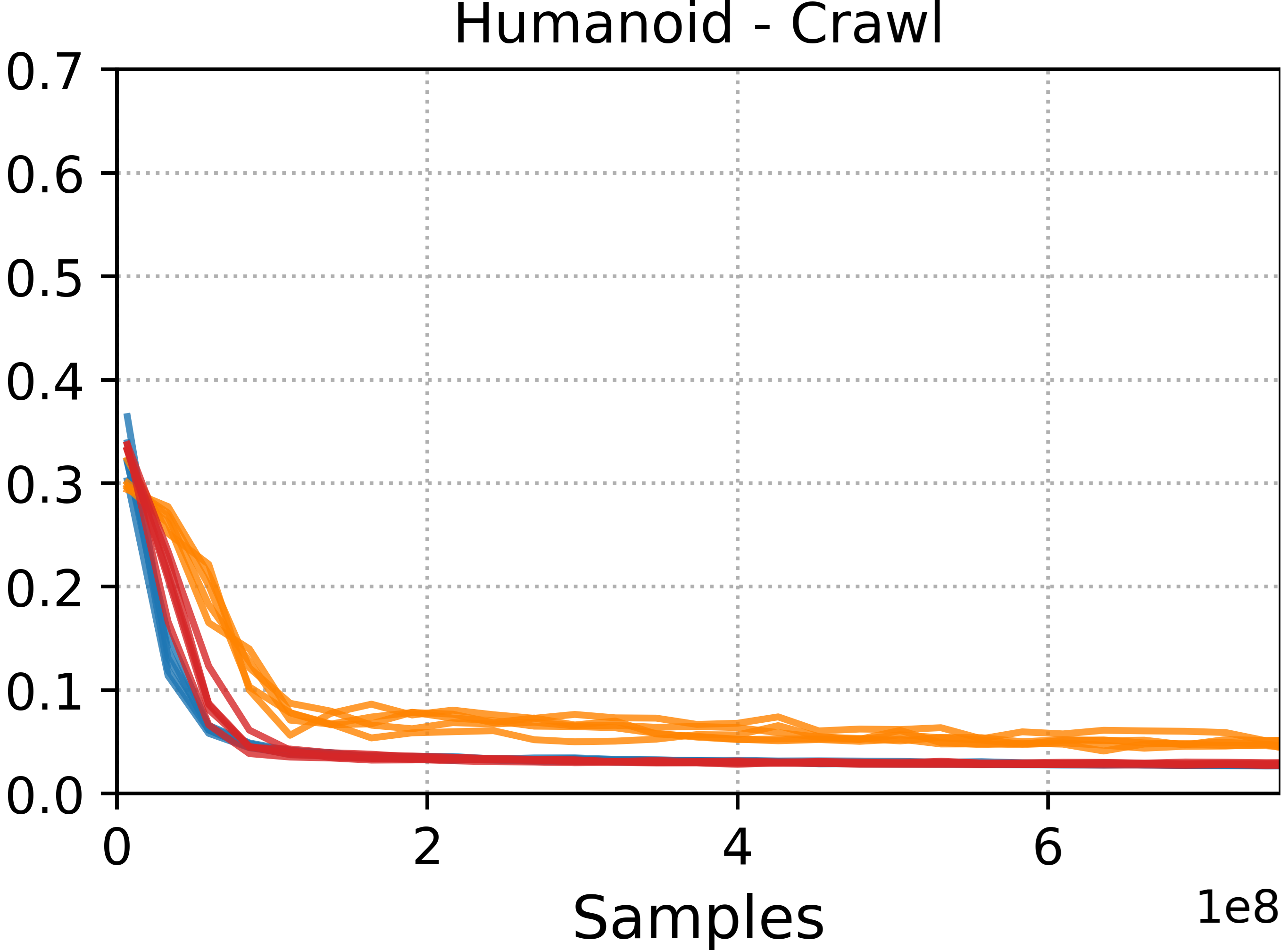}}
    \subfigure{\includegraphics[height=0.17\linewidth]{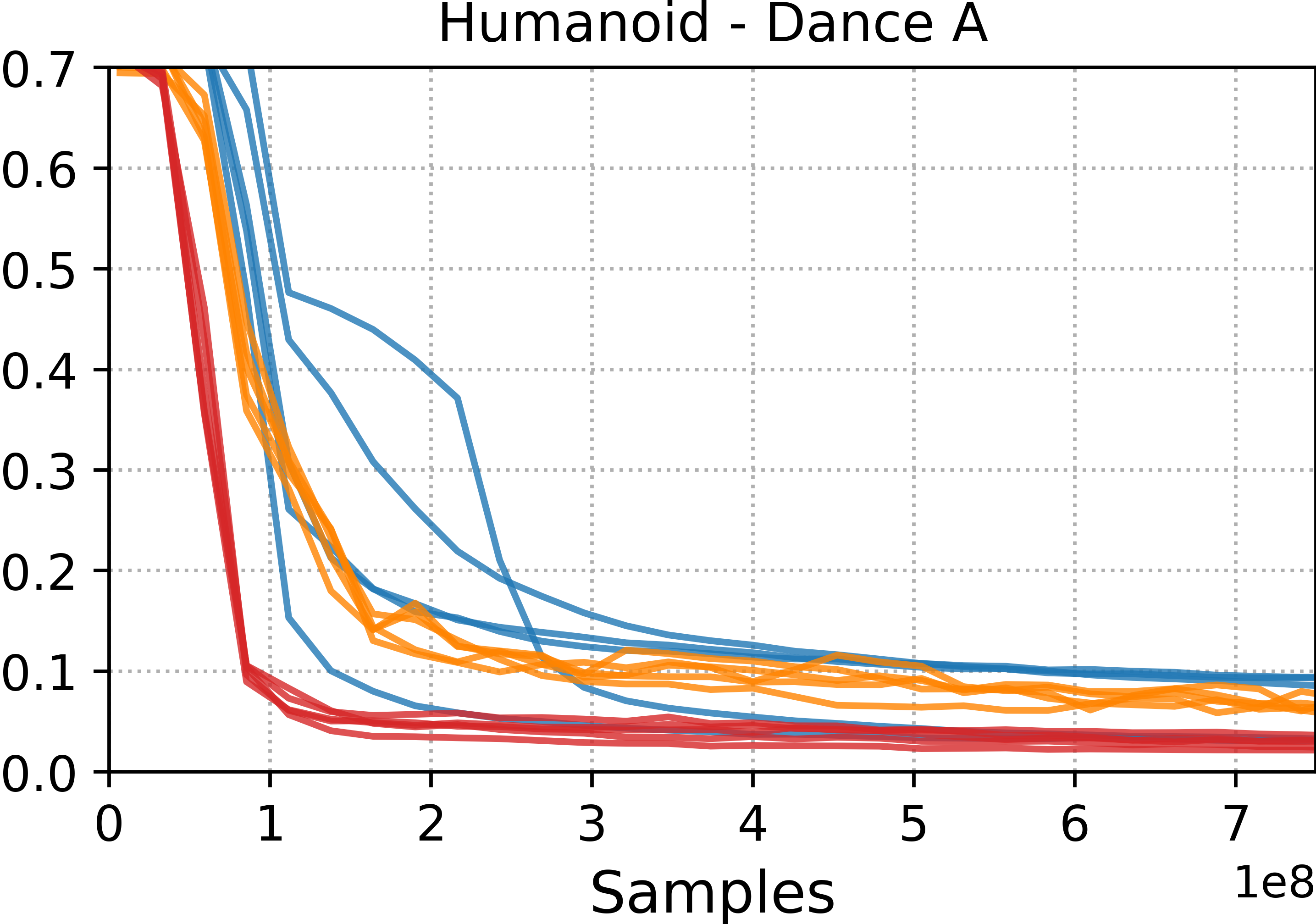}}\\
    \vspace{-0.2cm}
    \subfigure{\includegraphics[height=0.17\linewidth]{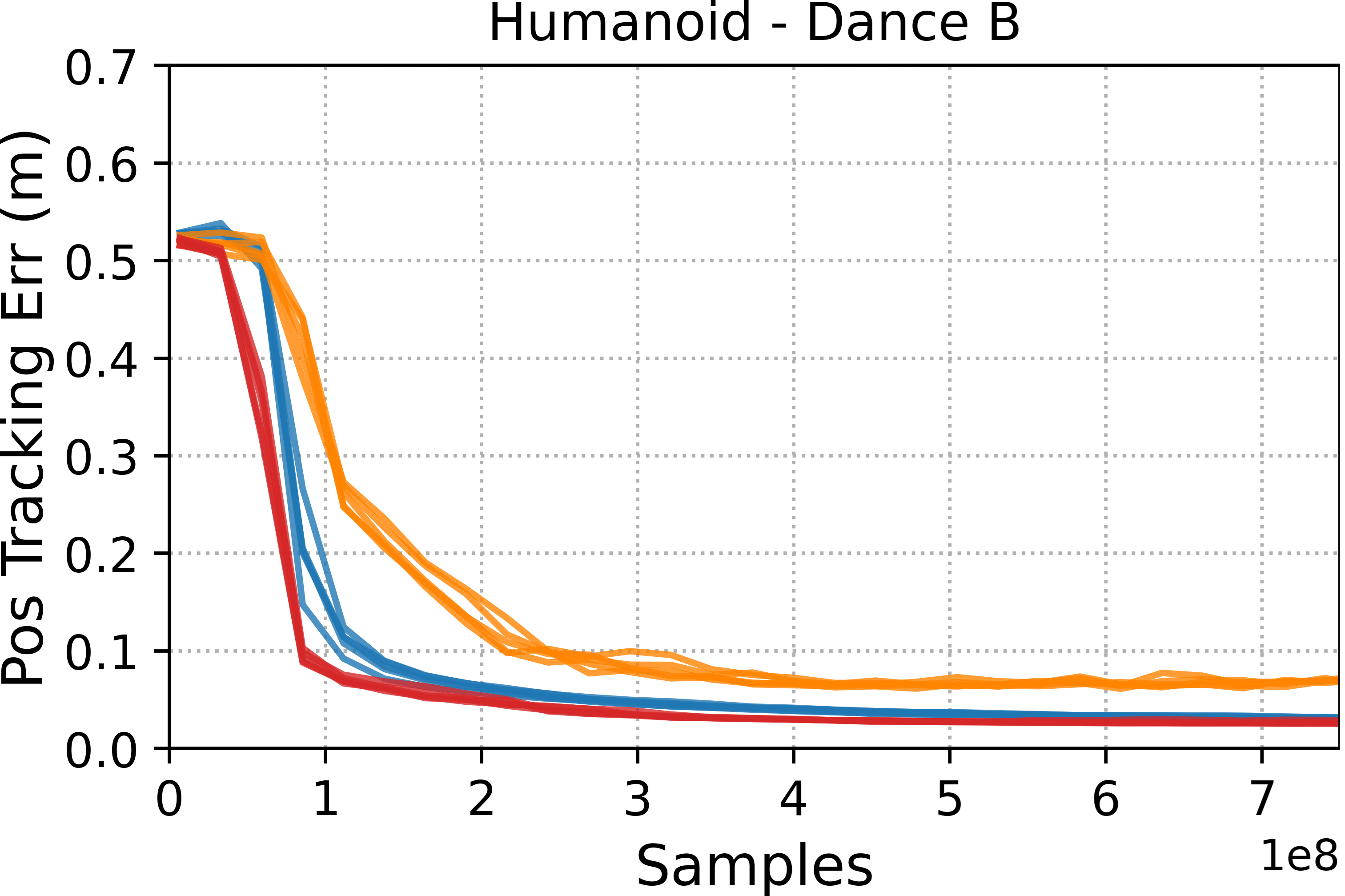}}
    \subfigure{\includegraphics[height=0.17\linewidth]{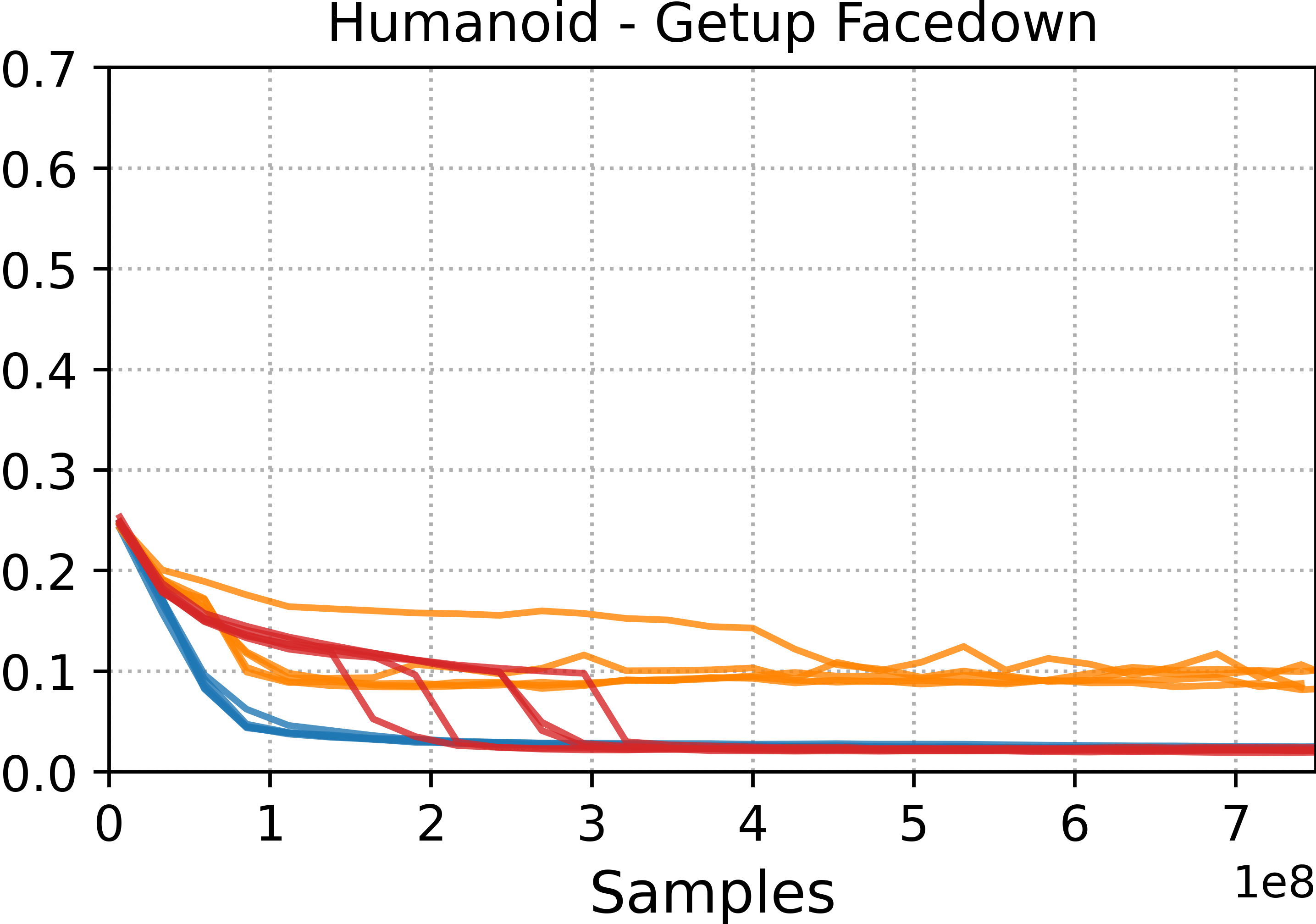}}
    \subfigure{\includegraphics[height=0.17\linewidth]{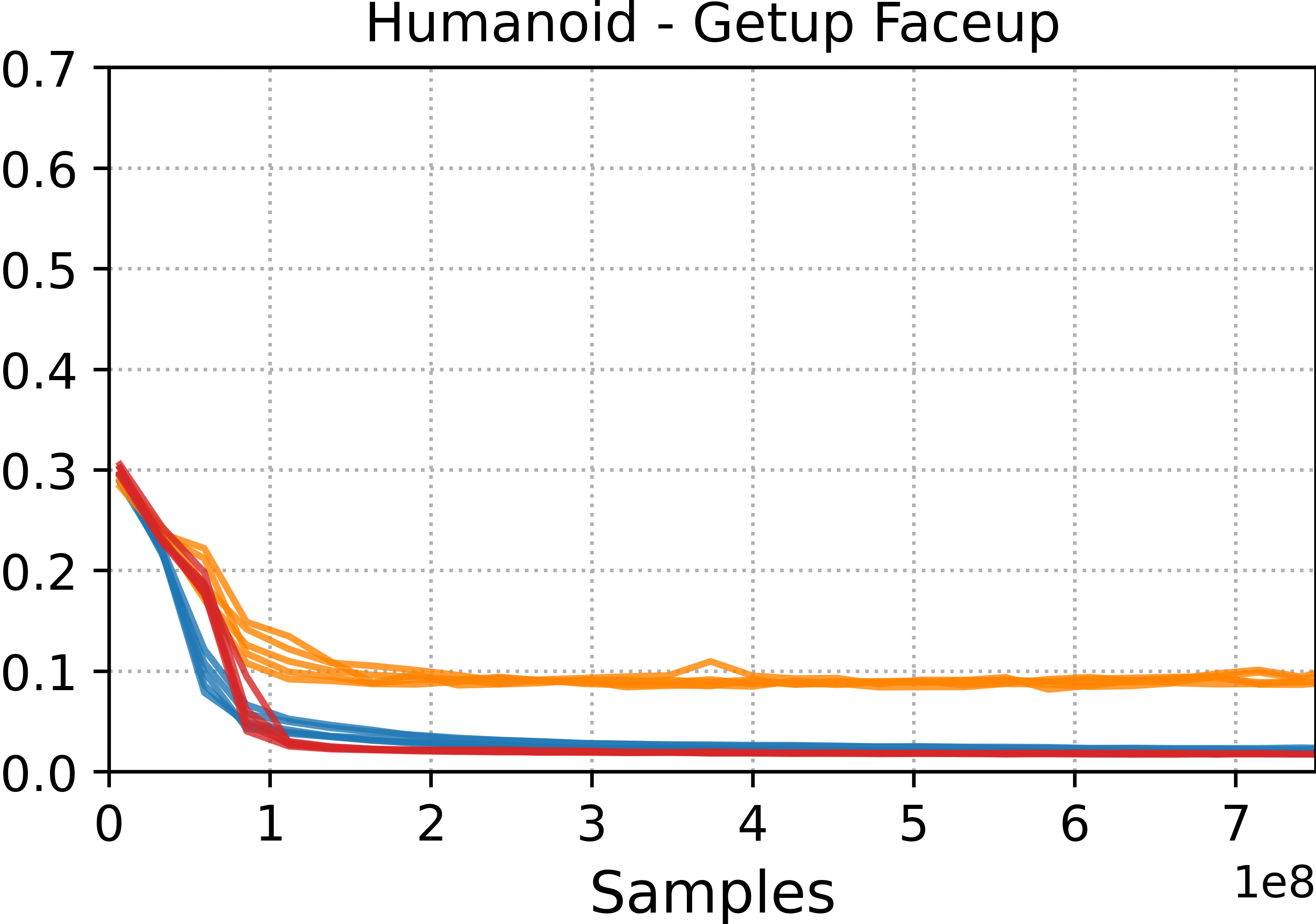}} 
    \subfigure{\includegraphics[height=0.17\linewidth]{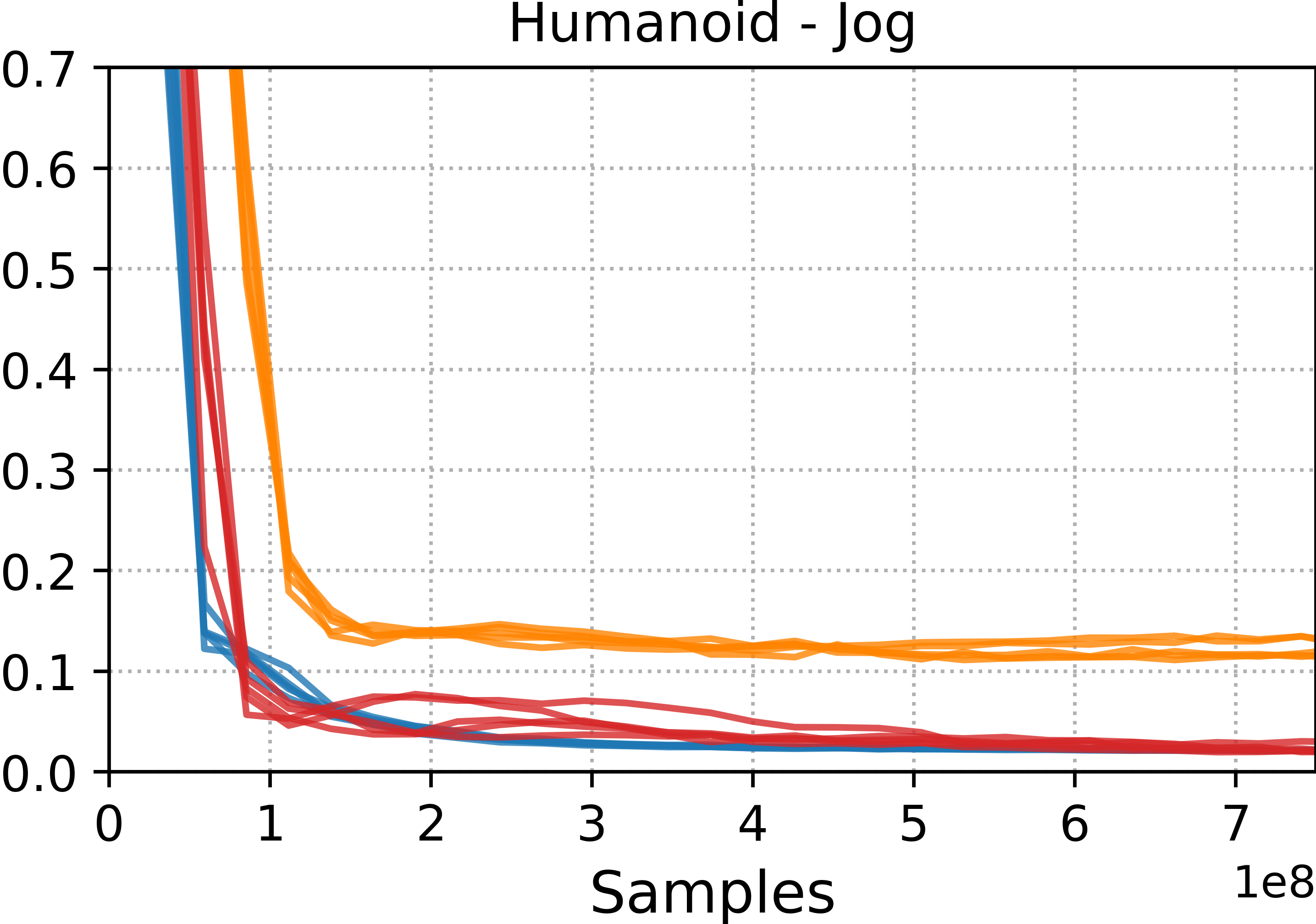}}\\
    \vspace{-0.2cm}
    \subfigure{\includegraphics[height=0.17\linewidth]{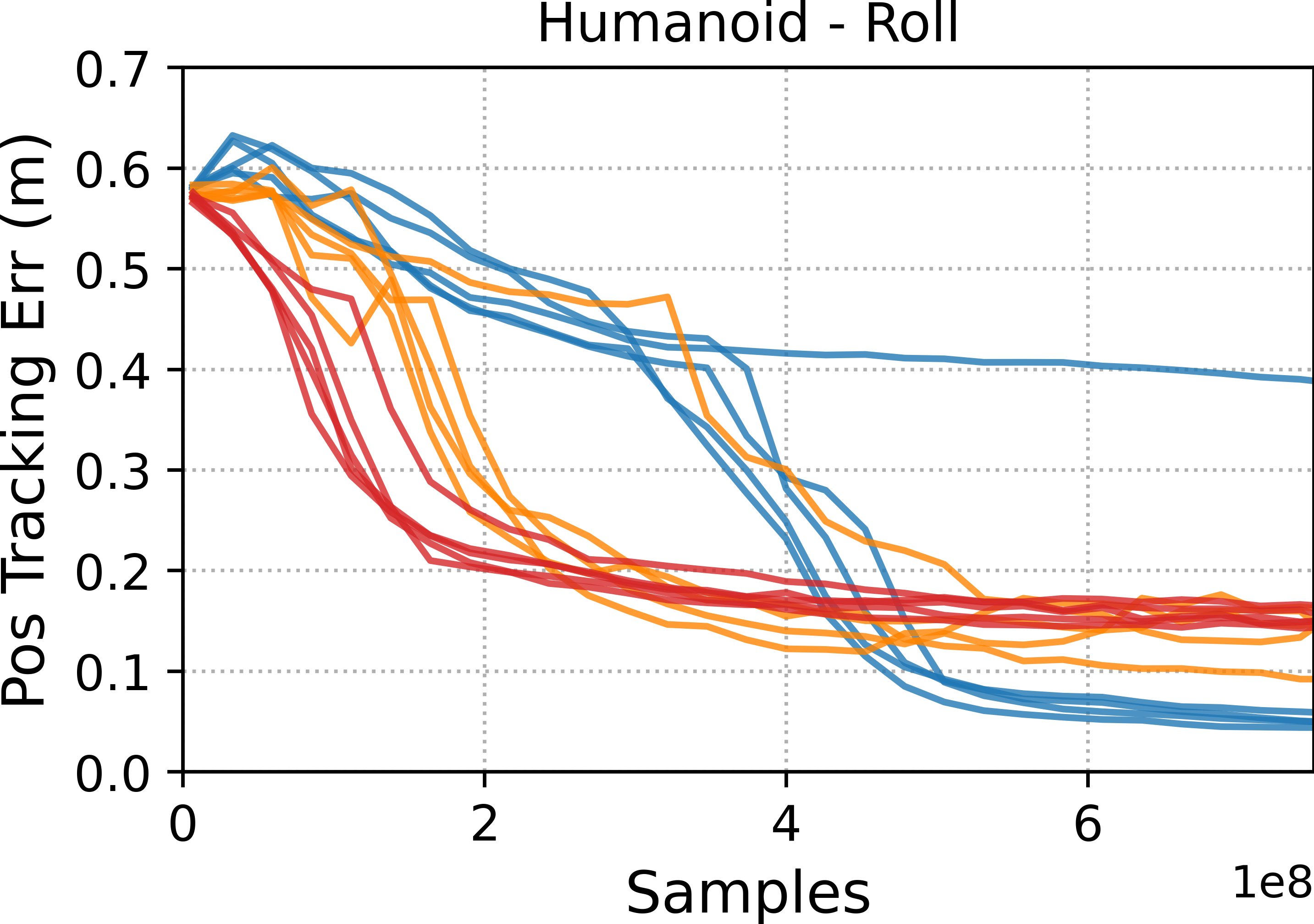}}
    \subfigure{\includegraphics[height=0.17\linewidth]{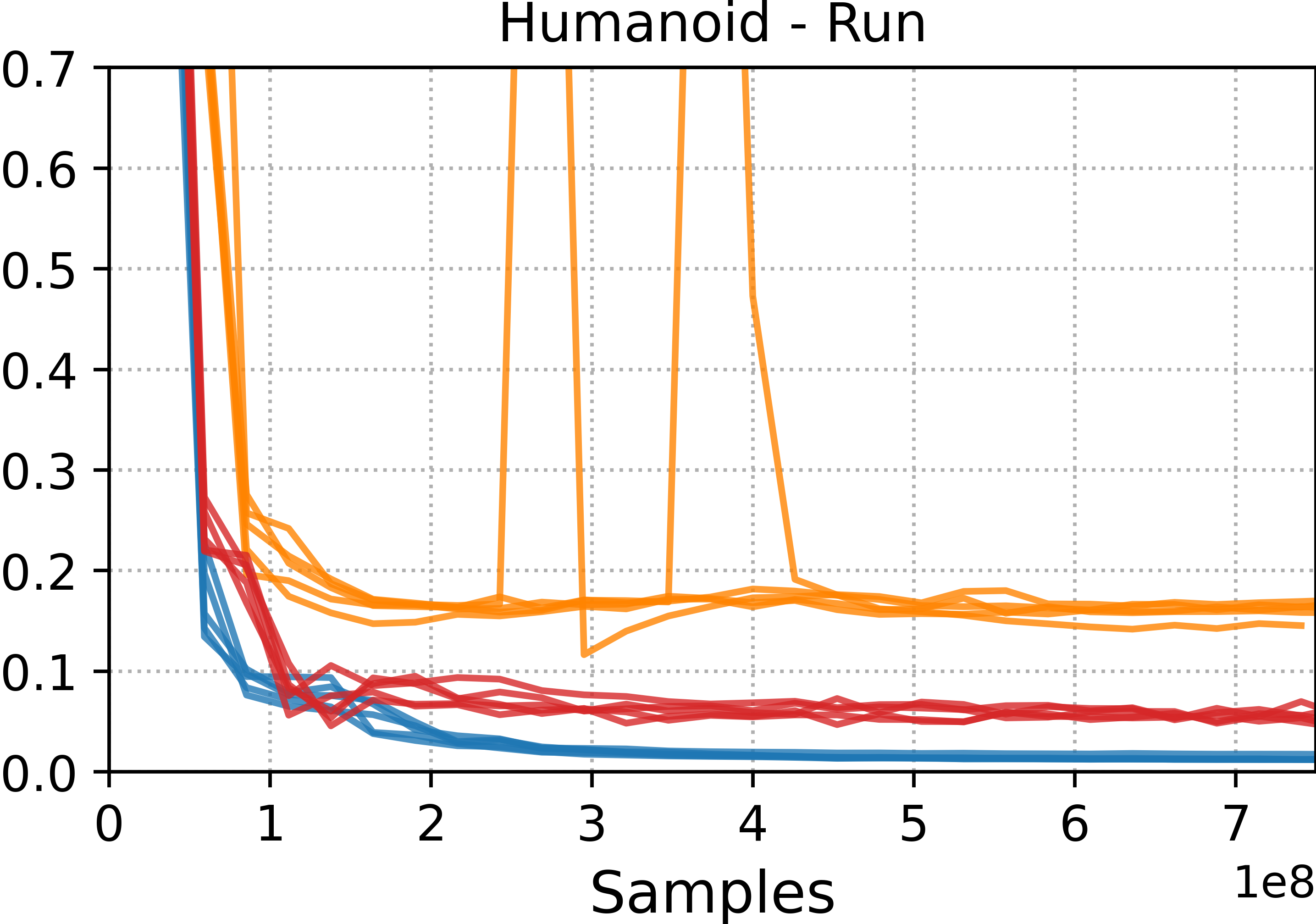}}
    \subfigure{\includegraphics[height=0.17\linewidth]{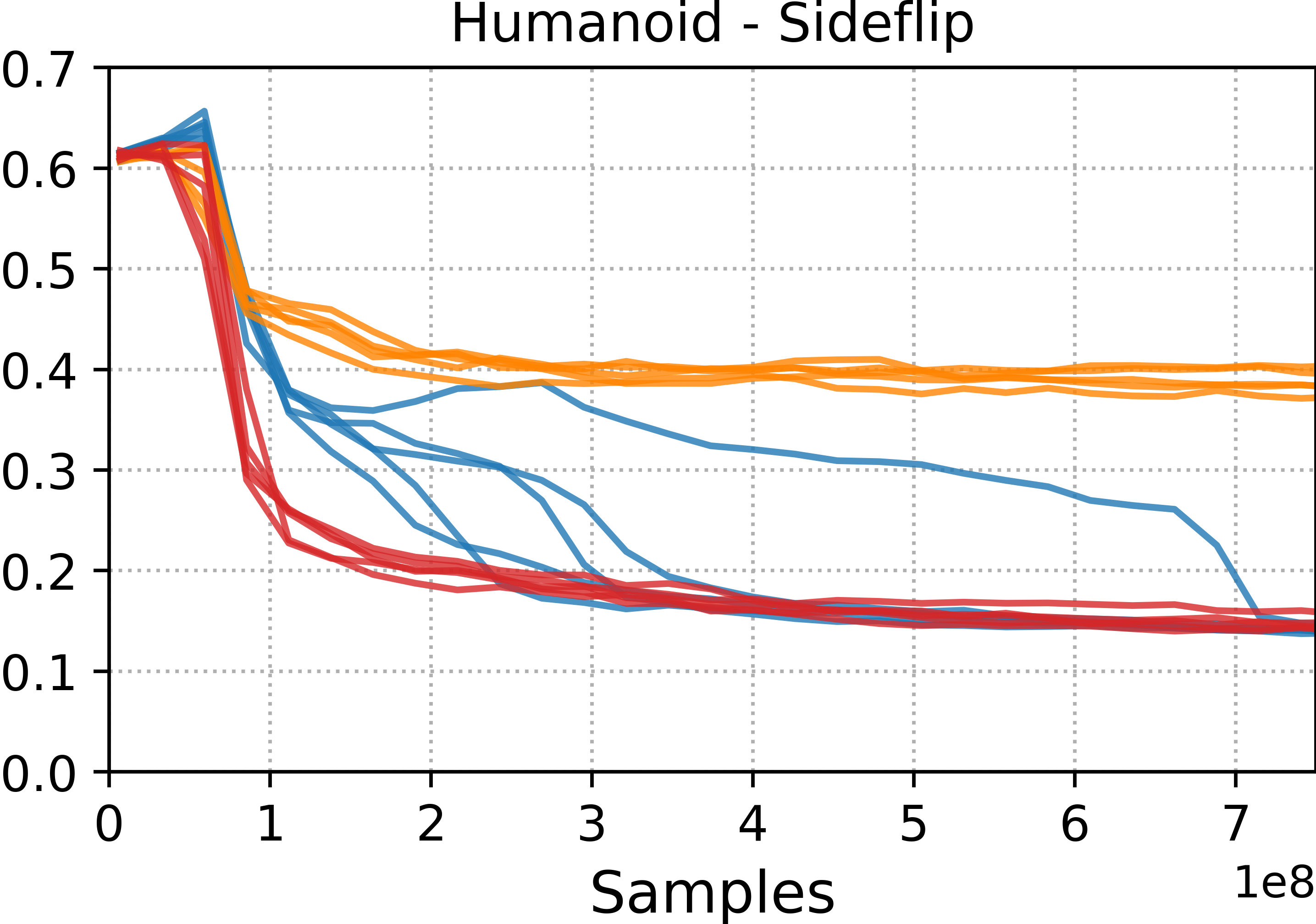}}
    \subfigure{\includegraphics[height=0.17\linewidth]{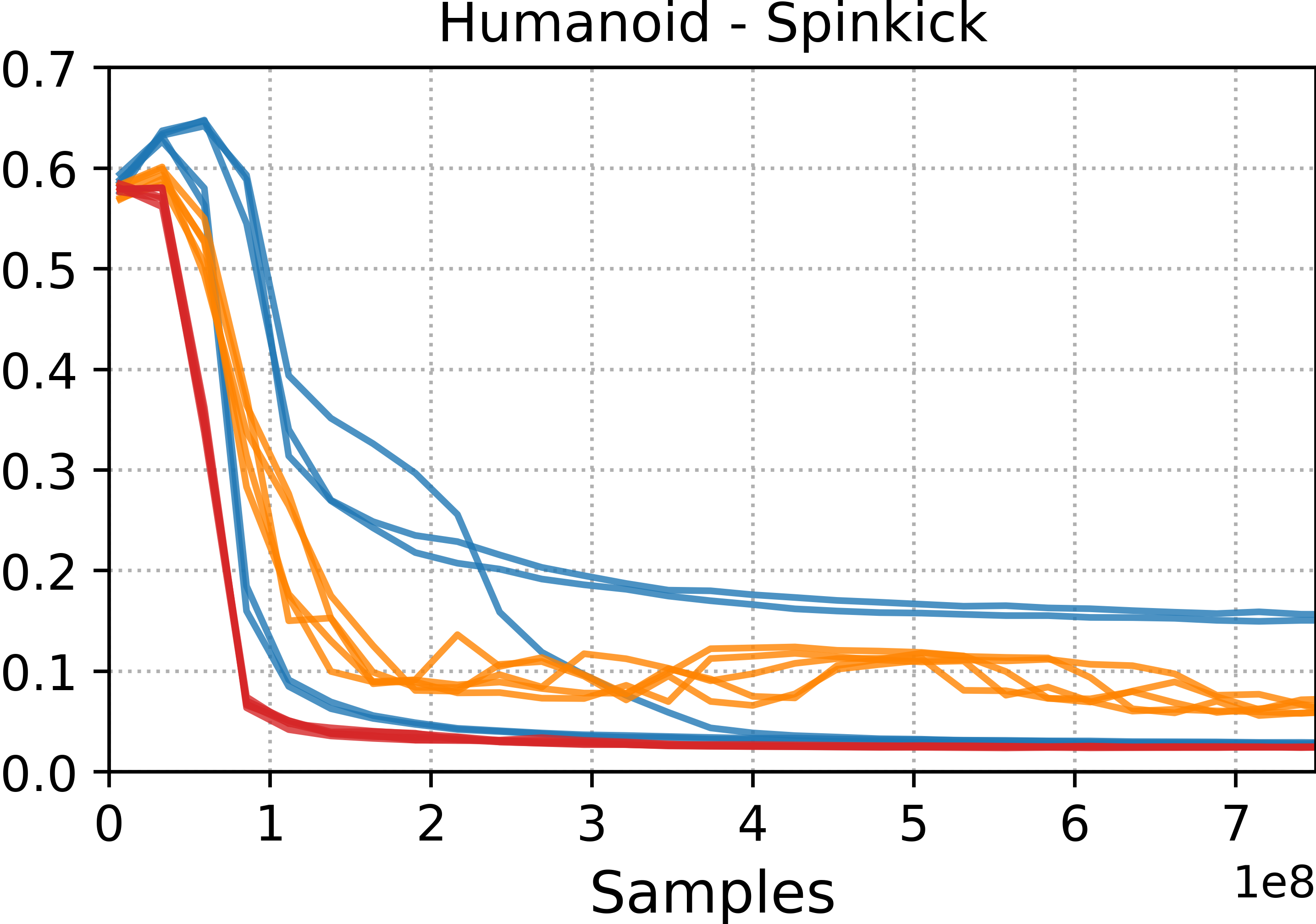}}\\
    \vspace{-0.1cm}
    \subfigure{\includegraphics[width=0.35\linewidth]{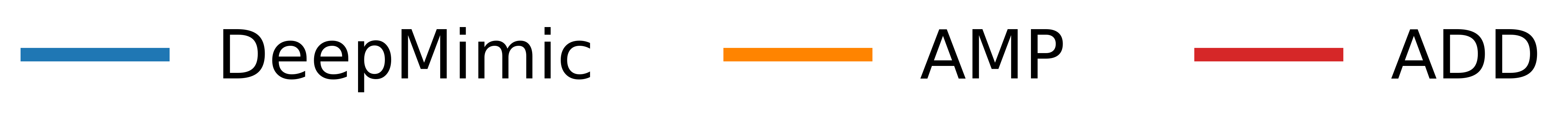}}\\
    \vspace{-0.5cm}
\caption{Learning curves comparing the tracking performance with the simulated humanoid character trained with DeepMimic, AMP, and ADD. Five training runs initialized with different random seeds are shown for each method. In order to better compare methods under similar settings, policies are trained \textbf{without pose-error termination}. The standard configuration for tracking-based methods, such as DeepMimic and ADD, utilizes pose-error termination, which tends to produce better performance and more consistent results across training runs.}
\label{fig:learning_curves_motion_tracking}
\end{figure}

During training, When using the TensorBoard logger during training, a TensorBoard \code{events} file will be saved the same output directory as the log file. The log can be viewed with:
\begin{lstlisting}[basicstyle=\ttfamily\small,backgroundcolor=\color{gray!20!white}]
tensorboard --logdir=output/ --port=6006 --bind_all
\end{lstlisting}
In addition to visualizing training statistics with the runtime loggers, output log \code{.txt} file can also be visualized using the plotting script \code{tools/plot\_log/plot\_log.py}. Examples of learning curves generated by \code{plot\_log.py} are shown in Figure~\ref{fig:learning_curves_motion_tracking}.

\section{Motion Data}

Most of the methods implemented in MimicKit utilize motion data to guide the training process. Example motion clips are provided in \code{data/motions/}. The \code{motion\_file} field in the environment configuration file can be used to specify the reference motion clip used for training and testing. In addition to imitating individual motion clips, \code{motion\_file} can also specify a dataset file, located in \code{data/datasets/}, which will train a model to imitate a dataset containing multiple motion clips.

The \code{view\_motion} environment can be used to visualize motion clips:
\begin{lstlisting}[basicstyle=\ttfamily\small,backgroundcolor=\color{gray!20!white}]
python mimickit/run.py --mode test --arg_file args/view_motion_humanoid_args.txt \
  --visualize true
\end{lstlisting}
Motion clips are represented by the \texttt{Motion} class implemented in \code{mimickit/anim/motion.py}. Each motion clip is stored in a \code{.pkl} file. Each frame in a motion specifies the pose of the character according to
\code{[root position (3D), root rotation (3D), joint rotations]},
where 3D rotations are specified using 3D exponential maps \citep{ExpMapGrassia1998}. Joint rotations are recorded in the order that the joints are specified in the \code{.xml} file (i.e. depth-first traversal of the kinematic tree). For example, in the case of the Humanoid character \code{data/assets/humanoid.xml}, each frame is represented as:
\vspace{0.2cm}

\begin{minipage}{0.3\textwidth}
\begin{enumerate}
    \item \code{root position} (3D)
    \item \code{root rotation} (3D)
    \item \code{abdomen} (3D)
    \item \code{neck} (3D)
    \item \code{right\_shoulder} (3D)
    \item \code{right\_elbow} (1D)
    \item \code{left\_shoulder} (3D)
    \item \code{left\_elbow} (1D)
    \item \code{right\_hip} (3D)
    \item \code{right\_knee} (1D)
    \item \code{right\_ankle} (3D)
    \item \code{left\_hip} (3D)
    \item \code{left\_knee} (1D)
    \item \code{left\_ankle} (3D)
\end{enumerate}
\hfill
\end{minipage}
\begin{minipage}{0.6\textwidth}
    \flushright 
    \includegraphics[width=1\textwidth]{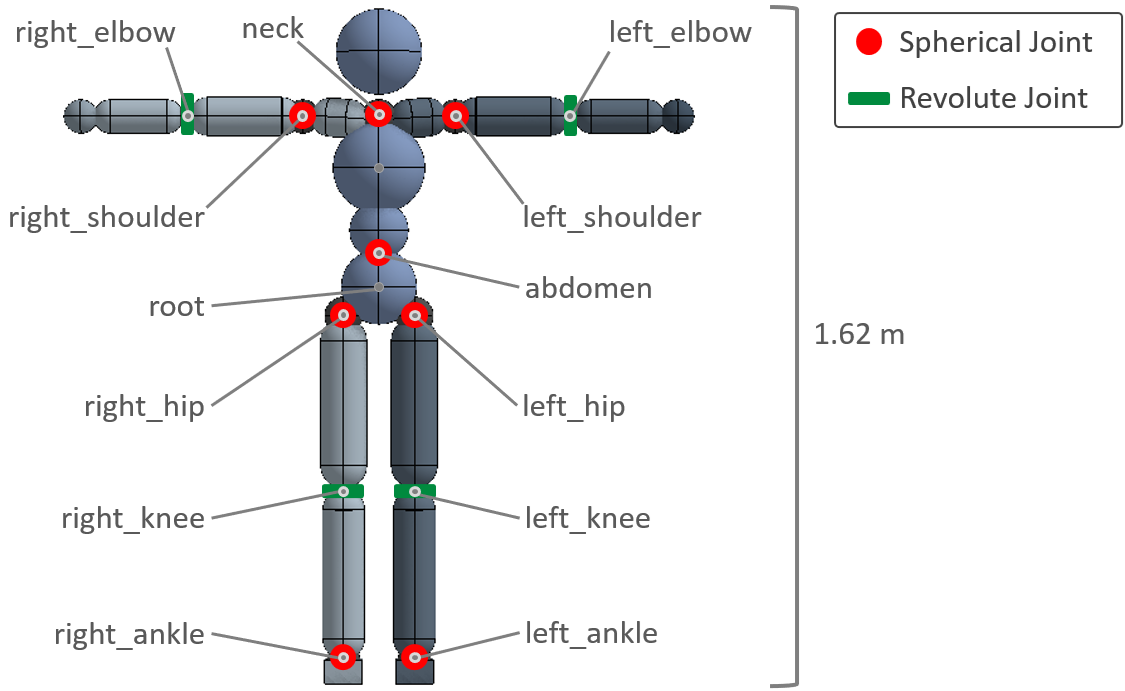}
    \captionof{figure}{Simulated Humanoid character.}
    \label{fig:humanoid}
\end{minipage}

\vspace{0.3cm}
\noindent The rotations of 3D joints are represented using 3D exponential maps, and the rotations of 1D joints are represented using 1D rotation angles.

\vspace{0.2cm}
\section{Experiments}

To evaluate the framework's effectiveness to reproduce diverse naturalistic motions, we apply the methods implemented in MimicKit on motion imitation tasks with a diverse suite of motions, ranging from common everyday behaviors, such as walking and running, to highly dynamic and athletic behaviors, such as acrobatics and martial arts. Our experiments assess both the quantitative tracking performance of the learned controllers and the qualitative fidelity of the resulting motions.

\subsection{Motion Imitation}
All experiments are conducted using the IsaacGym physics simulator. Each task involves a simulated humanoid character trained to imitate reference motion clips recorded via motion capture of live actors. We compare policies trained using three representative algorithms implemented within MimicKit: DeepMimic~\cite{2018-TOG-deepMimic}, AMP~\cite{2021-TOG-AMP}, and ADD~\cite{zhang2025ADD}. Separate policies are trained for each motion clip.
In order to compare different methods under similar settings, we disable pose error termination used in \citet{2018-TOG-deepMimic} during training, which terminates an episode if the character’s pose deviates significantly from the reference. Pose error termination is not applicable to distribution matching techniques such as AMP, where the policy is not synchronized with the reference motion. During training and evaluation, early termination is triggered only when the character makes undesired contact with the ground.

\begin{figure}[t!]
	\centering
    \subfigure[Humanoid - Backflip]{\includegraphics[width=0.495\columnwidth]{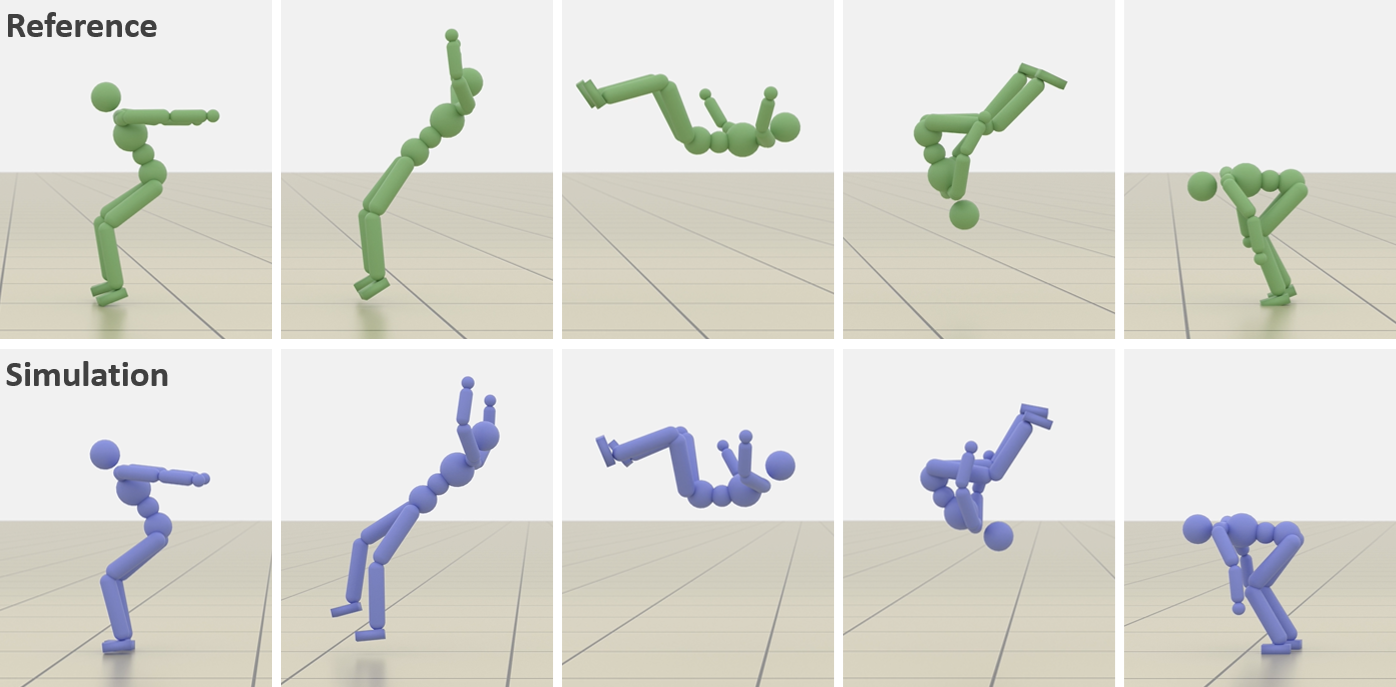}}
    \hfill
    \subfigure[Humanoid - Roll]{\includegraphics[width=0.495\columnwidth]{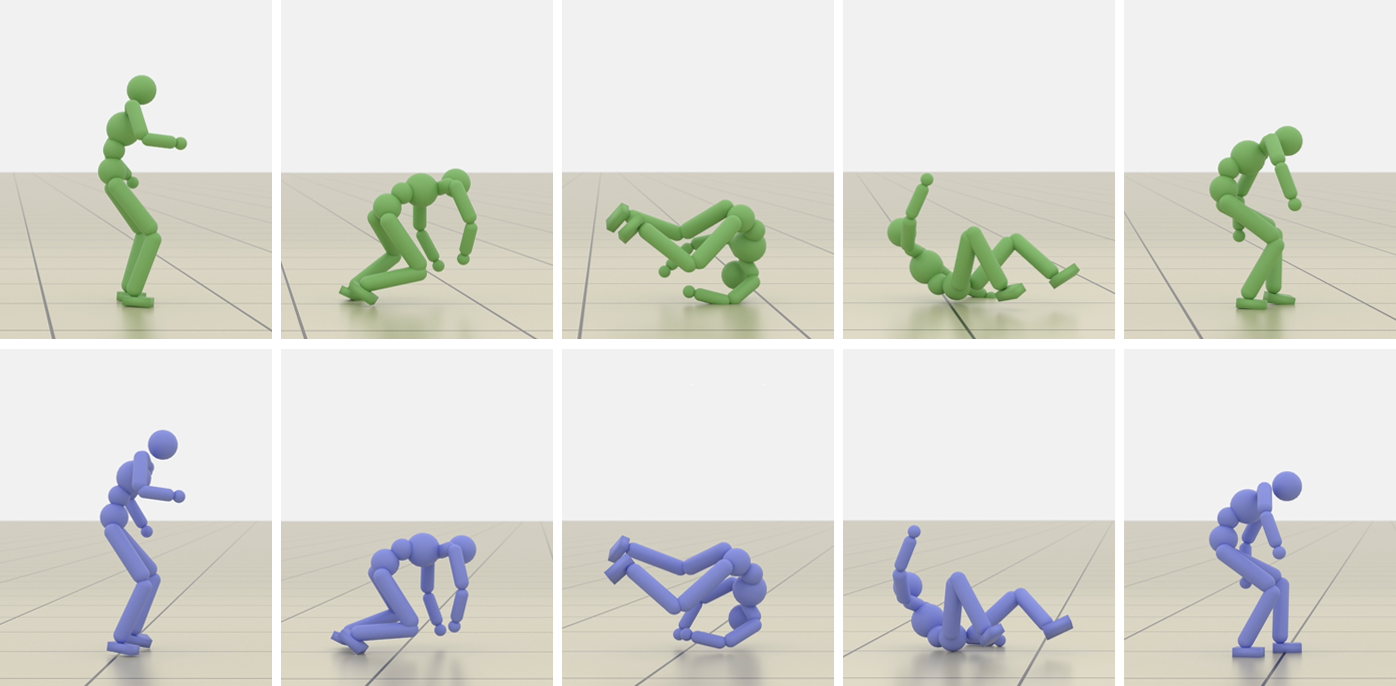}}\\
    \subfigure[SMPL - Spin]{\includegraphics[width=0.495\columnwidth]{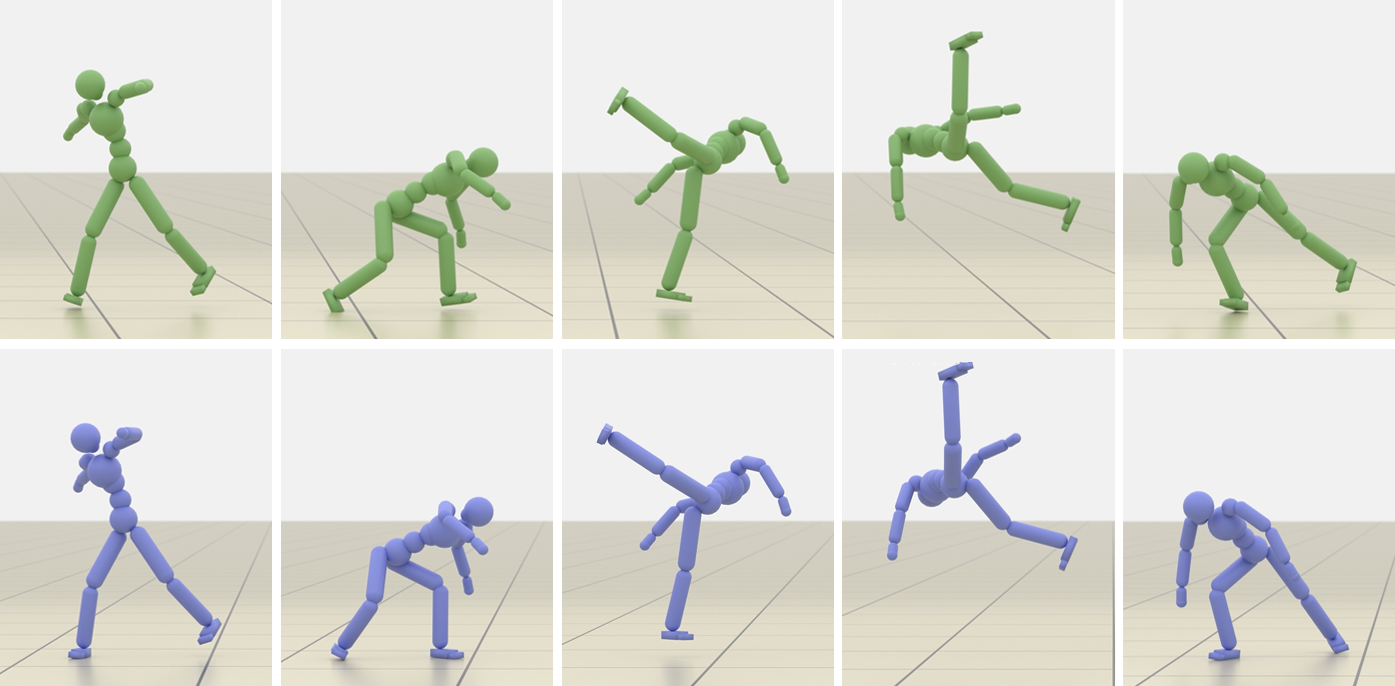}}
    \hfill
    \subfigure[SMPL - Getup Facedown]{\includegraphics[width=0.495\columnwidth]{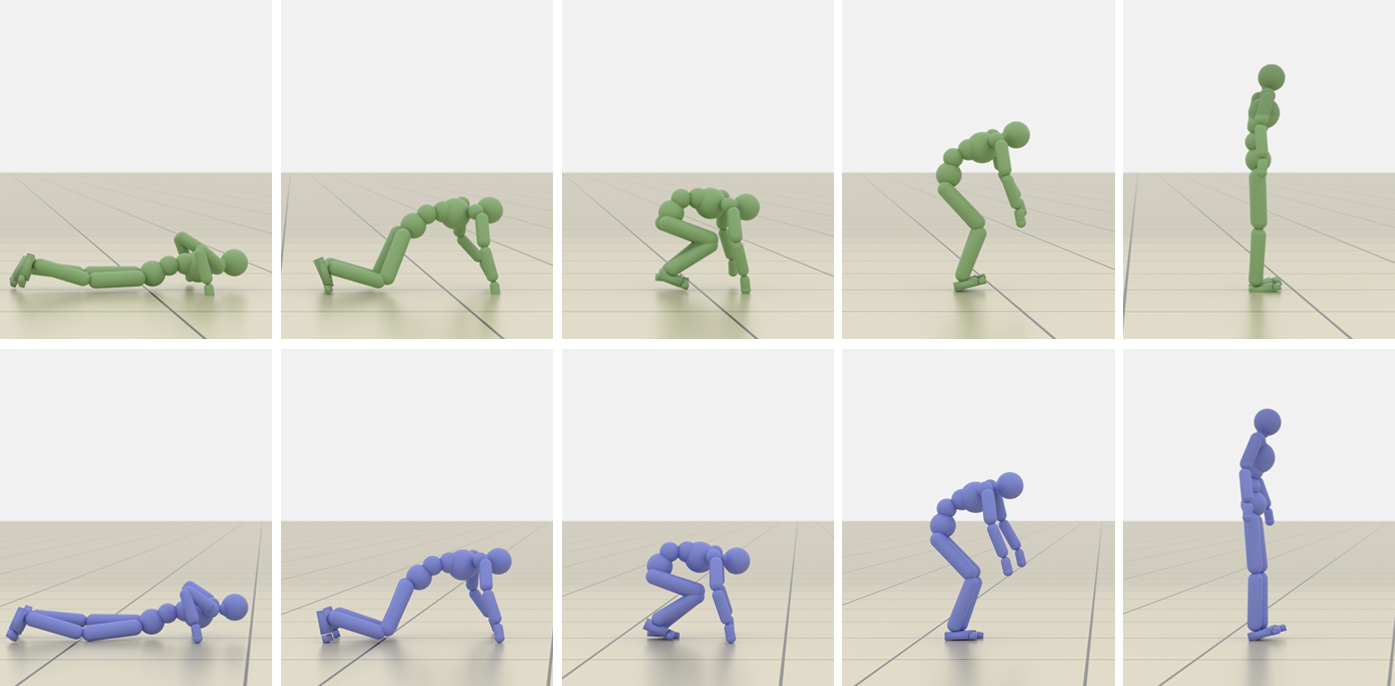}}\\
    \subfigure[G1 - Run]{\includegraphics[width=0.495\columnwidth]{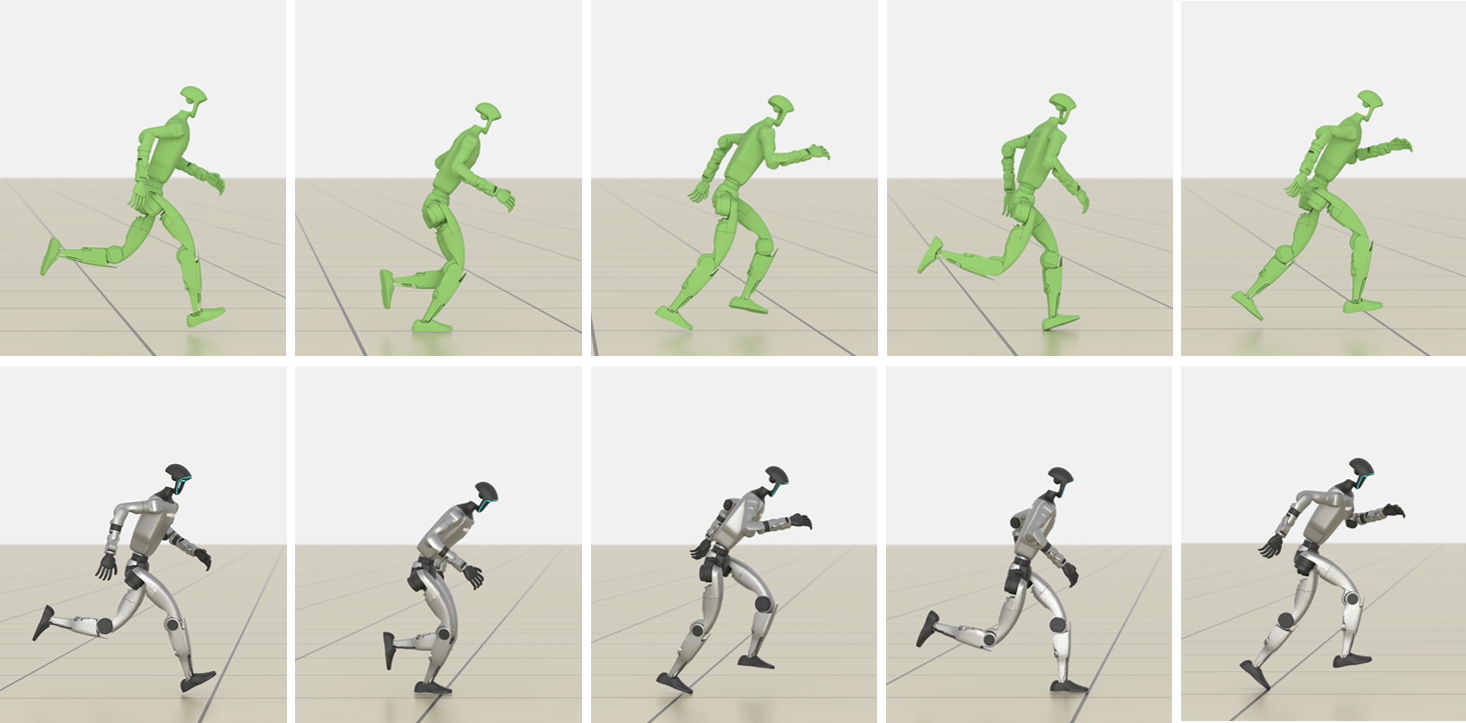}}
    \hfill
    \subfigure[G1 - Cartwheel]{\includegraphics[width=0.495\columnwidth]{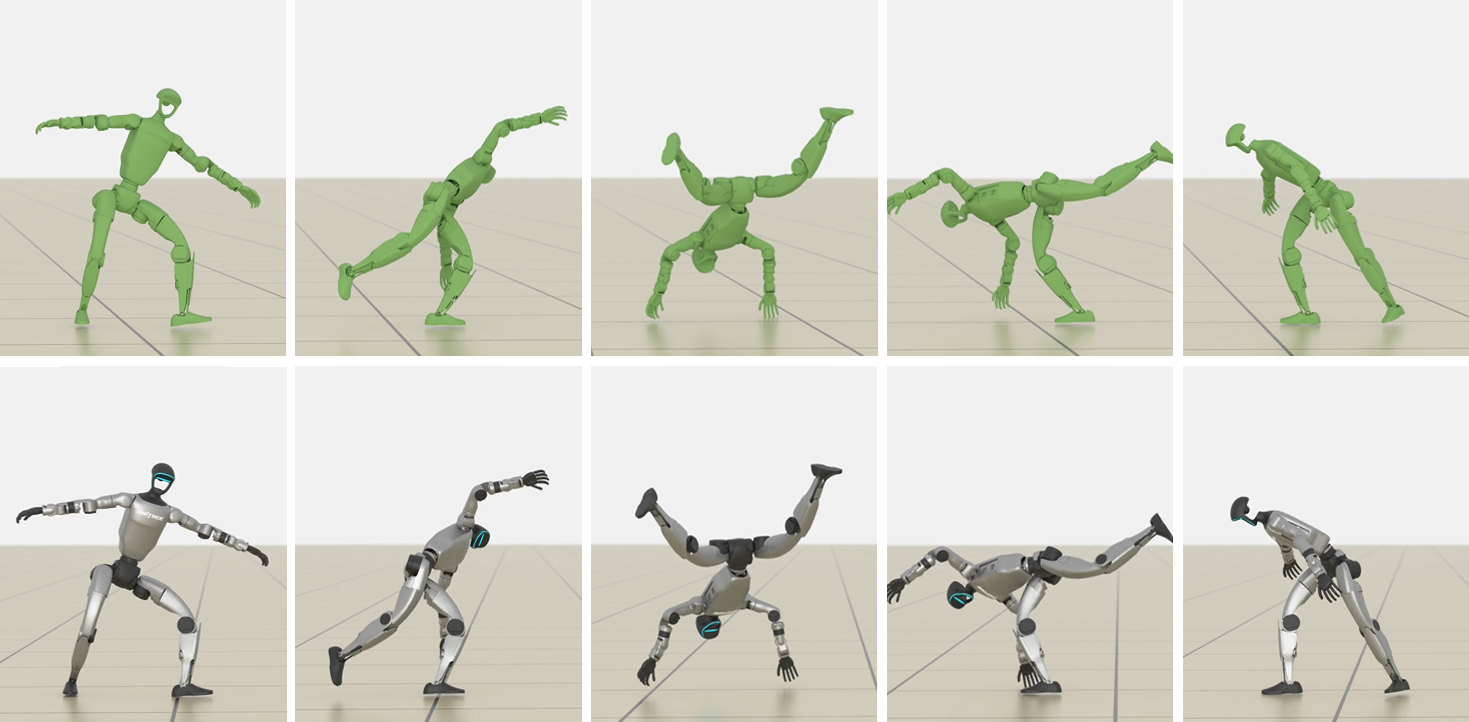}}\\
    \subfigure[Go2 - Trot]{\includegraphics[width=0.495\columnwidth]{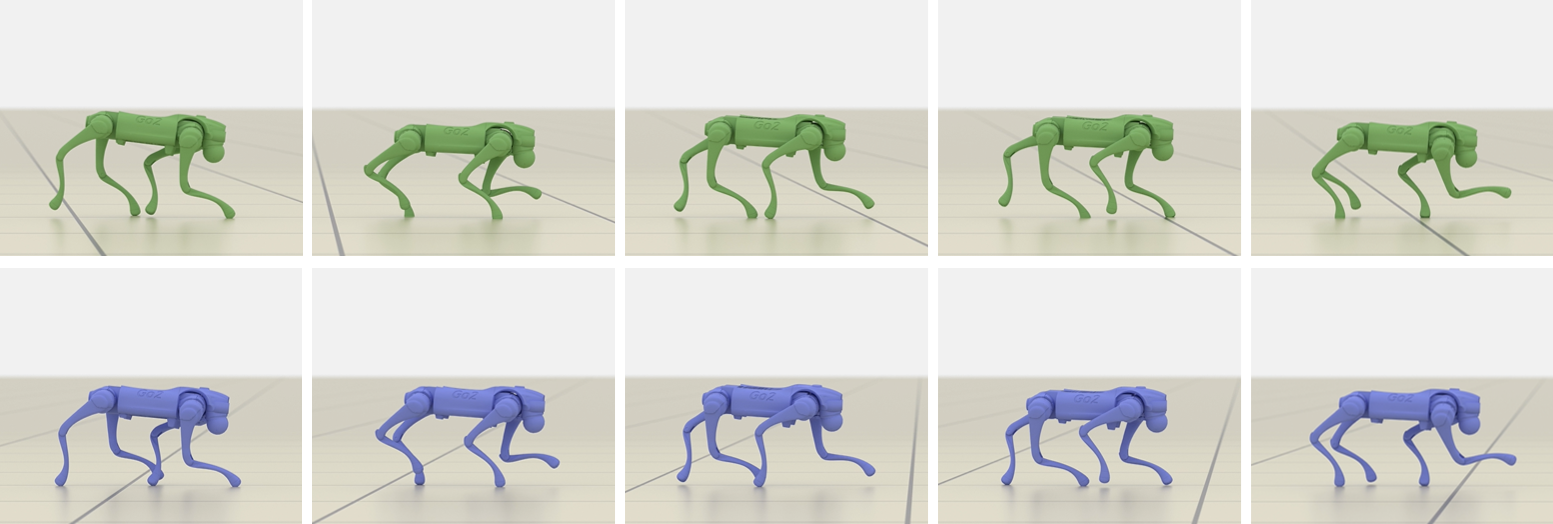}}
    \hfill
    \subfigure[Go2 - Canter]{\includegraphics[width=0.495\columnwidth]{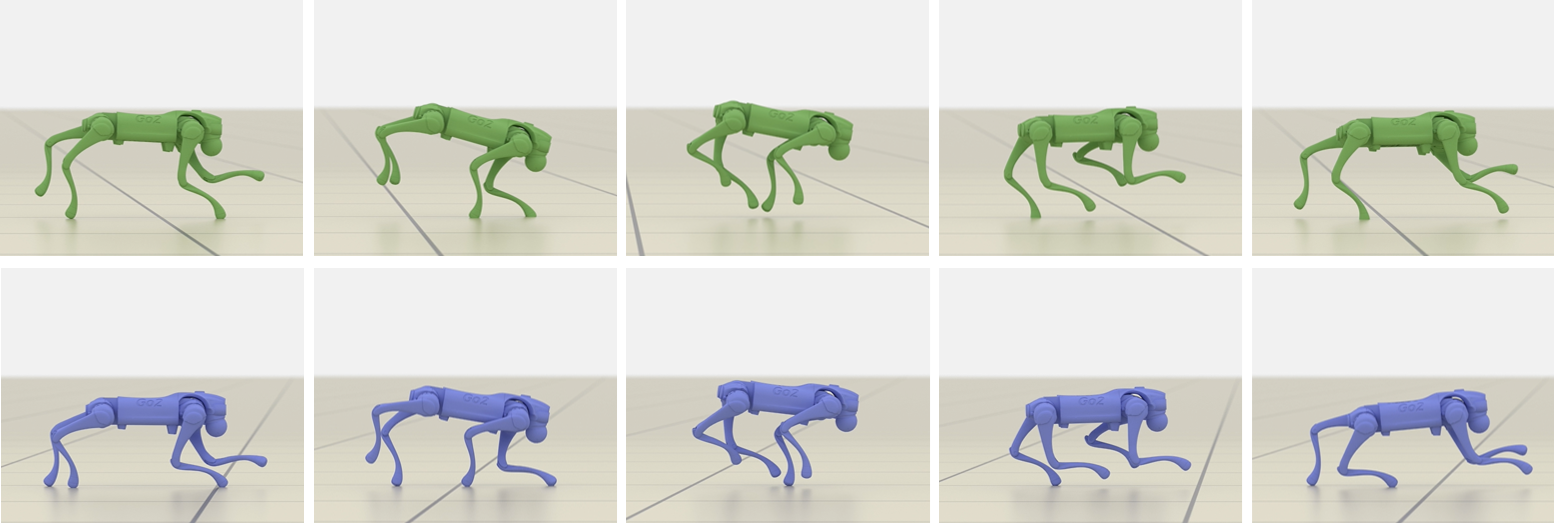}}\\
    \caption{Snapshots of physically simulated characters performing skills learned by imitating motion data recorded from real-life actors. The methods implemented in MimicKit can be applied to train policies for a diverse cast of simulated characters and skills.}
    \label{fig:filmstrips_tracking}
\end{figure}

\FloatBarrier
Snapshots of the behaviors learned by policies trained using various methods implemented in MimicKit are shown in Figure~\ref{fig:filmstrips_tracking}. Our framework is able to effectively train policies for a wide range of challenging and highly dynamics behaviors with a diverse cast of simulated characters, including a humanoid character modeled after the SMPL body model \citep{SMPL2015}, a Unitree G1 humanoid robot, and a Unitree Go2 quadrupedal robot. Despite the significant differences in morphology among these character, the same underlying training framework can be applied with minimal modifications, highlighting the modularity and generality of our system.

\begin{table}[t!]
\renewcommand{\arraystretch}{1.1} 
\centering
\begin{tabular}{|>{\centering\arraybackslash}m{1.4cm}|>{\centering\arraybackslash}m{1.5cm}|>{\centering\arraybackslash}m{1.7cm}|>{\centering\arraybackslash}m{1.7cm}|>{\centering\arraybackslash}m{1.7cm}|>{\centering\arraybackslash}m{1.7cm}|>{\centering\arraybackslash}m{1.7cm}|>{\centering\arraybackslash}m{1.7cm}|}
\hline
\textbf{Motion} & \textbf{Length} & \multicolumn{3}{c|}{\textbf{Position Tracking Error [m]}} & \multicolumn{3}{c|}{\textbf{DoF Velocity Tracking Error [rad/s]}} \\ \hline
               &                     & \textbf{AMP} & \textbf{DeepMimic} & \textbf{ADD} & \textbf{AMP} & \textbf{DeepMimic} & \textbf{ADD)} \\ \hline
Run            & 0.80s & $0.163^{\pm 0.008}$ & \cellcolor{blue!10}\(\mathbf{0.013^{\pm 0.002}}\)  & $0.165 ^{\pm 0.017}$ & $2.811 ^{\pm 0.048}$ & $0.584 ^{\pm 0.054}$ & \cellcolor{blue!10}\(\mathbf{0.478 ^{\pm 0.007}}\) \\ \hline
Jog            & 0.83s & $0.120^{\pm 0.007}$ & \cellcolor{blue!10}\(\mathbf{0.021^{\pm 0.000}}\) & $0.024^{\pm 0.004}$ & $2.017 ^{\pm 0.052}$ & $0.575 ^{\pm 0.007}$ & \cellcolor{blue!10}\(\mathbf{0.507 ^{\pm 0.010}}\) \\ \hline
Sideflip       & 2.44s & $0.387^{\pm 0.011}$ & \cellcolor{blue!10}$\mathbf{0.138^{\pm 0.004}}$ & $0.145^{\pm 0.006}$ & $2.276 ^{\pm 0.014}$ & \cellcolor{blue!10}$\mathbf{1.118 ^{\pm 0.034}}$ & $1.350 ^{\pm 0.049}$ \\ \hline
Crawl          & 2.93s & $0.050^{\pm 0.006}$ & \cellcolor{blue!10}$\mathbf{0.027^{\pm 0.000}}$ & $0.028^{\pm 0.002}$ & $0.646 ^{\pm 0.089}$ & $0.430 ^{\pm 0.006}$ & \cellcolor{blue!10}$\mathbf{0.283 ^{\pm 0.002}}$ \\ \hline
Roll           & 2.00s & $0.141^{\pm 0.031}$ & \cellcolor{blue!10}$\mathbf{0.115^{\pm 0.132}}$ & $0.152^{\pm 0.005}$ & $1.576 ^{\pm 0.318}$ & \cellcolor{blue!10}$\mathbf{0.994 ^{\pm 0.051}}$ & $1.330 ^{\pm 0.101}$ \\ \hline
Getup Facedown & 3.03s & $0.096^{\pm 0.018}$ & $0.023^{\pm 0.001}$ & \cellcolor{blue!10}$\mathbf{0.022^{\pm 0.001}}$ & $0.838 ^{\pm 0.029}$ & $0.433 ^{\pm 0.008}$ & \cellcolor{blue!10}$\mathbf{0.325 ^{\pm 0.005}}$ \\ \hline
Spinkick       & 1.28s & $0.064^{\pm 0.010}$ & $0.078^{\pm 0.062}$ & \cellcolor{blue!10}$\mathbf{0.025^{\pm 0.000}}$ & $1.453 ^{\pm 0.327}$ & $1.222 ^{\pm 0.233}$ & \cellcolor{blue!10}$\mathbf{0.774 ^{\pm 0.007}}$ \\ \hline
Cartwheel      & 2.71s & $0.076^{\pm 0.006}$ & $0.144^{\pm 0.153}$ & \cellcolor{blue!10}$\mathbf{0.017^{\pm 0.000}}$ & $0.722 ^{\pm 0.020}$ & $0.659 ^{\pm 0.160}$ & \cellcolor{blue!10}$\mathbf{0.317 ^{\pm 0.002}}$ \\ \hline
Backflip       & 1.75s & $0.267^{\pm 0.015}$ & $0.111^{\pm 0.054}$ & \cellcolor{blue!10}$\mathbf{0.062^{\pm 0.001}}$ & $2.243 ^{\pm 0.113}$ & $1.103 ^{\pm 0.024}$ & \cellcolor{blue!10}$\mathbf{0.878 ^{\pm 0.013}}$ \\ \hline
Dance A        & 1.62s & $0.065^{\pm 0.009}$ & $0.065^{\pm 0.029}$ & \cellcolor{blue!10}$\mathbf{0.028^{\pm 0.007}}$ & $0.895 ^{\pm 0.108}$ & $0.830 ^{\pm 0.090}$ & \cellcolor{blue!10}$\mathbf{0.428 ^{\pm 0.014}}$ \\ \hline
Walk           & 0.96s & $0.132^{\pm 0.021}$ & $0.009^{\pm 0.001}$ & \cellcolor{blue!10}$\mathbf{0.009^{\pm 0.001}}$ & $1.394 ^{\pm 0.123}$ & $0.286 ^{\pm 0.005}$ & \cellcolor{blue!10}$\mathbf{0.213 ^{\pm 0.003}}$ \\ \hline
\end{tabular}\\
\vspace{0.25cm}
\caption{Motion tracking performance of the Humanoid character trained using AMP, DeepMimic, and ADD. Position (Eq. \ref{Eq:pos_tracking_err}) and DoF Velocity tracking errors are averaged across 5 models initialized with different random seeds. For each model, errors are calculated using 4096 test episodes. Motion tracking methods, such as DeepMimic and ADD, are able to more accurately reproduce a given reference motion compared to distribution-matching methods, such as AMP.}
\label{tab:humanoid_motion_comp}
\end{table}

To evaluate the performance of each policy, we measure the position tracking error $e^\mathrm{pos}_t$, and DoF velocity tracking error $e^\mathrm{vel}_t$, which provides an indicator of motion smoothness. The position tracking error $e^\mathrm{pos}_t$ measures the difference in the global root position and relative joint positions between the simulated character and the reference motion:
\begin{equation}
\label{Eq:pos_tracking_err}
e^\mathrm{pos}_t = \frac{1}{N^\mathrm{joint} + 1} \left( \sum\limits_{j \in \mathrm{joints}} \left|\left|(\hat{\rvx}^j_t - \hat{\rvx}^\mathrm{root}_t) - (\rvx^j_t - \rvx^\mathrm{root}_t)\right|\right|_2 + \left|\left|\hat{\rvx}^\mathrm{root}_t - \rvx^\mathrm{root}_t\right|\right|_2 \right).
\end{equation}
Here, $\rvx^j_t$ and $\hat{\rvx}^j_t$ represent the 3D Cartesian position of joint $j$ from the simulated character and the reference motion, respectively. $N^\mathrm{joint}$ denotes the number of joints in the character. The DoF velocity tracking error measures the differences in local angular velocities of each joint between the simulated character and the reference motion:
\begin{equation}
e^\mathrm{vel}_t = \frac{1}{N^\mathrm{joint} + 1} \sum\limits_{j \in \mathrm{joints}} \left|\left|\hat{\dot{\rvq}}^j_t - \dot{\rvq}^j_t\right|\right|_2,
\end{equation}
where $\dot{\rvq}^j_t$ and $\hat{\dot{\rvq}}^j_t$ represent the local angular velocity of joint $j$ from the simulated character and the reference motion.

Table~\ref{tab:humanoid_motion_comp} summarizes performance of the various methods. Performance statistics for each method are calculated across 5 models initialized with different random seeds. AMP exhibits poor tracking performance, since the policies are trained using a general distribution-matching objective. However, qualitatively AMP can still be effective at reproducing the general behaviors of a reference motion, despite not precisely tracking the motion clip. Motion tracking methods, such as DeepMimic and ADD are able to accurately track a wide variety of reference motions. However, there are important distinctions in the consistency of the results across training runs. Since DeepMimic relies on a manually-designed reward function, it can be difficult to craft a general reward function that can effectively and consistently imitate a diverse variety of behaviors, in the absence of additional heuristics such as pose error termination. In contrast, ADD leverages a differential discriminator to automatically learn an adaptive reward function, which can lead to more consistent performance across diverse motions. However, we would like to note that when pose error termination is enabled during training, tracking accuracy and consistency across training runs generally improve substantially. Therefore, the default configuration for tracking-based methods generally incorporate pose error termination during training.

\section{Conclusion}
In this work, we introduced MimicKit, an open-source reinforcement learning framework for motion imitation and control. MimicKit a unifies a suite of motion imitation methods for training motion controllers within a modular and extensible framework. We hope MimicKit will facilitate reproducible research in motor skill learning, and provide a convenient platform to accelerate progress in learning-based methods for motion control.

\section*{Acknowledgments}

We would like to thank all of the collaborators who have made valuable contributions to the development of MimicKit (ordered alphabetically): Yuxuan Mu, Yi Shi, Michael Xu, Dun Yang, and Ziyu Zhang. This work was supported by NSERC (RGPIN-2015-04843), the National Research Council Canada (AI4D-166), and Sony Interactive Entertainment.

\bibliographystyle{ACM-Reference-Format}
\bibliography{MimicKit}

\end{document}